\documentclass[11pt,journal, onecolumn, draftcls]{IEEEtran}
\usepackage{amsmath}
\usepackage{epsfig}
\usepackage{amssymb}
\usepackage{graphicx}
\usepackage{hyperref}
\usepackage{color}
\usepackage{subfigure}
\usepackage{algorithmic}
\usepackage{algorithm}
\usepackage{psfrag}
\begin{document}
\title{Sum-Rate Maximization in Two-Way AF MIMO Relaying: Polynomial
Time Solutions to a Class of DC Programming Problems}
%\ninept
\author{Arash~Khabbazibasmenj,~\IEEEmembership{Student Member,~IEEE},
Florian~Roemer,~\IEEEmembership{Student Member,~IEEE}, \\ Sergiy
A.~Vorobyov\thanks{S.~A.~Vorobyov is the corresponding author.
A.~Khabbazibasmenj and S.~A.~Vorobyov are with the Department of
Electrical and Computer Engineering, University of Alberta,
Edmonton, AB, T6G~2V4 Canada; e-mail: \{{\tt khabbazi,
svorobyo\}@ualberta.ca}. F.~Roemer and M.~Haardt are with the
Communication Research Laboratory, Ilmenau University of
Technology, Ilmenau, 98693, Germany; \{{\tt florian.roemer,
martin.haardt\}@tu-ilmenau.de}.

This work is supported in parts by the Natural Science and
Engineering Research Council (NSERC) of Canada and the Carl Zeiss
Award, Germany. This work has been performed in the framework of
the European research project SAPHYRE, which is partly funded by
the European Union under its FP7 ICT Objective 1.1 - The Network
of the Future. The trip of the first co-author to Germany has been
also partly supported by the Graduate School on Mobile
Communications (GSMobicom), Ilmenau University of Technology,
which is partly funded by the Deutsche Forschungsgemeinschaft
(DFG).

Parts of this paper have been published at EUSIPCO~2010, Aalborg,
Denmark and have been submitted to
ICASSP~2012.},~\IEEEmembership{Senior Member,~IEEE}, and
Martin~Haardt,~\IEEEmembership{Senior Member,~IEEE}}
\vspace{-1cm}
\maketitle
\vspace{-1.5cm}
%\begin{center}
%\centerline{ EDICS: SAM-SAMC}
%\end{center}
%\vspace{-0.5cm}
\begin{abstract}
Sum-rate maximization in two-way amplify-and-forward (AF)
multiple-input multiple-output (MIMO) relaying belongs to the
class of difference-of-convex functions (DC) programming problems.
DC programming problems occur as well in other signal processing
applications and are typically solved using different
modifications of the branch-and-bound method. This method,
however, does not have any polynomial time complexity guarantees.
In this paper, we show that a class of DC programming problems, to
which the sum-rate maximization in two-way MIMO relaying belongs,
can be solved very efficiently in polynomial time, and develop two
algorithms. The objective function of the problem is represented
as a product of quadratic ratios and parameterized so that its
convex part (versus the concave part) contains only one (or two)
optimization variables. One of the algorithms is called
POlynomial-Time DC (POTDC) and is based on semi-definite
programming (SDP) relaxation, linearization, and an iterative
search over a single parameter. The other algorithm is called
RAte-maximization via Generalized EigenvectorS (RAGES) and is
based on the generalized eigenvectors method and an iterative
search over two (or one, in its approximate version) optimization
variables. We also derive an upper-bound for the optimal values of
the corresponding optimization problem and show by simulations
that this upper-bound can be achieved by both algorithms. The
proposed methods for maximizing the sum-rate in the two-way AF
MIMO relaying system are shown to be superior to other
state-of-the-art algorithms.
\end{abstract}

%\newpage

\section{Introduction}
Two-way relaying has recently attracted a significant research
interest due to its ability to overcome the drawback of
conventional one-way relaying, that is, the factor of 1/2 loss in
the rate \cite{Rankov}, \cite{Oecht}. Moreover, two-way relaying
can be viewed as a certain form of network coding \cite{Asw} which
allows to reduce the number of time slots used for the
transmission in one-way relaying by relaxing the requirement of
`orthogonal/non-interfering' transmissions between the terminals
and the relay \cite{ZhangCui}. Specifically, simultaneous
transmissions by the terminals to the relay on the same
frequencies are allowed in the first time slot, while a combined
signal is broadcasted by the relay in the second time slot. In
contract to the one-way relaying case, the rate-optimal strategy
for two-way relaying is in general unknown \cite{Rong}. However,
some efficient strategies have been developed. Depending on the
ability of the relay to regenerate/decode the signals from the
terminals, several two-way transmission protocols have been
introduced and studied. The regenerative relay adopts the
decode-and-forward protocol and performs the decoding process at
the relay \cite{Hamm}, while the non-regenerative relay typically
adopts a form of amplify-and-forward (AF) protocol and does not
perform decoding at the relay, but amplifies and possibly
beamforms or precodes the signals to retransmit them back to the
terminals \cite{Rong}, \cite{Joung}, \cite{Amah}. The advantages
of the latter are a smaller delay in the transmission and lower
hardware complexity of the relay.

In this paper, we consider the AF two-way relaying system with two
terminals equipped with a single antenna and one relay with
multiple antennas. The task is to find the relay transmit strategy
that maximizes the sum rate of both terminals. This is a basic
model which can be extended in many ways. The significant
advantage of considering this basic model is that the
corresponding capacity region is discussed in the existing
literature in \cite{ZhangCui}. It enables us to concentrate on the
mathematical issues of the corresponding optimization problem
which are of significant and ubiquitous interest.

We show that the optimization problem of finding the relay
amplification matrix for the considered AF two-way relaying system
is equivalent to finding the maximum of the product of quadratic
ratios under a quadratic power constraint on the available power
at the relay. Such a problem belongs to the class of the so-called
difference-of-convex functions (DC) programming problems. It is
worth stressing that DC programming problems are very common in
signal processing and, in particular, signal processing for
communications. For example, the robust adaptive beamforming for
the general-rank (distributed source) signal model with a positive
semi-definite constraint can be shown to belong to the class of DC
programming problems \cite{GerChen}, \cite{ArashMe}. Specifically,
the constraint in the corresponding optimization problem is the
difference of two weighted norm functions. The power control for
wireless cellular systems is also a DC programming problem when
the the rate is used as a utility function \cite{KevinMe}.
Similarly, the dynamic spectrum management for digital subscriber
lines \cite{LeNgoc} as well as the problems of finding the
weighted sum-rate point, the proportional-fairness operating
point, and the max-min optimal point (egalitarian solution) for
the two-user multiple input single output (MISO) interference
channel \cite{JL10} are all DC programming problems. The typical
approach for solving such problems is the use of various
modifications of the branch-and-bound method
\cite{JL10}-\cite{Tuy2} that is an efficient global optimization
method. The branch-and-bound method is known to work well
especially for the case of monotonic functions, i.e., the case
which is typically encountered in signal processing and, in
particular, signal processing for communications. However, it does
not have any worst-case polynomial complexity guarantees, which
significantly limits or essentially prohibits its applicability in
practical communication systems. Thus, methods with guaranteed
polynomial-time complexity that can solve different types of DC
programming problems are of a fundamental importance.

In the last decade, a significant progress has occurred in the
application of optimization theory in signal processing and
communications. Some of those results are relevant for the
considered problem of maximizing constrained product of quadratic
ratios \cite{Vorobyov}-\cite{DeMaio}. The worst-case-based robust
adaptive beamforming problem is known to belong to the class of
second-order cone (SOC) programming problems \cite{Vorobyov}
largely due to the fact that the output
signal-plus-interference-to-noise ratio (SINR) of adaptive
beamforming is unchanged when the beamforming vector undergoes an
arbitrary phase rotation. This allows to simplify the single
worst-case distortionless response constraint of the optimization
problem into the form of a SOC constraint. The situation is
significantly more complicated in the case of multiple constraints
of the same type as the constraint in \cite{Vorobyov} when a
single rotation of the beamforming vector is not sufficient to
satisfy all constraints simultaneously. This situation is
successfully addressed in \cite{Sid} by considering the
semi-definite programming (SDP) relaxation technique. The SDP
relaxation technique has been then further developed and studied
in, for example, \cite{Yonina}, \cite{DeMaio} and other works.
Interestingly, the work \cite{DeMaio} considers the fractional
quadratically constrained quadratic programming (QCQP) problem
that is closest to the one addressed in this paper with the
significant difference though that the objective in \cite{DeMaio}
contains only a single quadratic ratio that simplifies the problem
dramatically.

In this paper, we develop polynomial time algorithms for finding
the globally optimal solution of a class of non-convex DC
programming problems, e.g., the maximization of a product of
quadratic ratios under a quadratic constraint. This problem
precisely corresponds to the sum-rate maximization in two-way AF
MIMO relaying. Our algorithms use such parameterizations of the
objective function that its convex part (versus the concave part)
contains only one (or two) optimization variables. One of the
proposed algorithms is named POlynomial-Time DC (POTDC) and is
based on semi-definite programming (SDP) relaxation,
linearization, and an iterative search over a single
parameter.\footnote{Some preliminary results on the POTDC
algorithm have been submitted to ICASSP'12 \cite{ArashMeFlorian}.}
The POTDC algorithm is rigorous and finds the global maximum of
the considered problem. Indeed, the solution given by this
algorithm coincides with the newly developed upper-bound for the
optimal value of the problem. The other algorithm is called
RAte-maximization via Generalized EigenvectorS (RAGES) and is
based on the generalized eigenvectors method and an iterative
search over two (or one, in its approximate version) optimization
variables.\footnote{Some preliminary results on the RAGES
algorithm have been presented in \cite{Florian}.} The RAGES
algorithm is somewhat heuristic in its approximate version, but
may enjoy a lower complexity.

The rest of the paper is organized as follows. The two-way AF MIMO
relaying system model is given in Section~II while the sum-rate
optimization problem for the corresponding system is formulated in
Section~III. The POTDC algorithm for solving the corresponding
sum-rate maximization is developed in Section~IV and an
upper-bound for the optimal value of the maximization problem is
found in Section~V. In Section~VI, the RAGES algorithm is
developed and investigated. Simulation results are reported in
Section~VII followed by the conclusions. This paper is
reproducible research \cite{Rep} and the software needed to
generate the simulation results will be provided to the IEEE
Xplore together with the paper upon its acceptance.

\section{System Model}
We consider a two-way relaying system with two single-antenna
terminals and an amplify-and-forward (AF) relay equipped with
$M_R$ antennas. Fig.~1 shows the system we study in the paper. In
the first transmission phase, both terminals transmit to the
relay. Assuming frequency-flat quasi-static block fading, the
received signal at the relay can be expressed as
\begin{align}
{\bf r} = {\bf h}_1^{(f)} \cdot x_1 + {\bf h}_2^{(f)} \cdot x_2 +
{\bf n}_{R}
\end{align}
where ${\bf h}_i^{(f)} = [h_{i,1}, \ldots, h_{i,M_R}]^T \in
\mathbb{C}^{M_R}$ represents the (forward) channel vector between
terminal $i$ and the relay, $x_i$ is the transmitted symbol from
terminal $i$, ${\bf n}_{R} \in \mathbb{C}^{M_R}$ denotes the
additive noise component at the relay, and $( \cdot )^T$ stands
for the transpose of a vector or a matrix. Let $P_{T,i} =
\mathbb{E}\{|x_i|^2\}$ be the average transmit power of terminal
$i$, ${\bf R}_{N,R} = \mathbb{E}\{{\bf n}_{R} \cdot {\bf n}_{R}^H
\}$ be the noise covariance matrix at the relay, $\mathbb{E}\{
\cdot \}$ denoting the mathematical expectation, and $( \cdot )^H$
standing for the Hermitian transpose of a vector or a matrix. For
the special case of white noise we have ${\bf R}_{N,R} = P_{N,R}
\cdot {\bf I}_{M_R}$ where $P_{N,R} = {\rm tr} ({\bf R}_{N,R}) /
M_R$ and ${\bf I}_{M_R}$ is the identity matrix of size $M_R
\times M_R$.

The relay amplifies the received signal by multiplying it with a
relay amplification matrix ${\bf G} \in \mathbb{C}^{M_R \times
M_R}$, i.e., it transmits the signal ${\bf \bar{r}} = {\bf G}
\cdot {\bf r}$. The transmit power used by the relay can be
expressed as
\begin{align}
\mathbb{E} \{\| {\bf{\bar{r}}} \|_2^2 \} & = \mathbb{E} \left\{
{\rm tr} \left\{ {\bf G} \cdot {\bf r} \cdot {\bf r}^H \cdot {\bf
G}^H \right\} \right\} \nonumber \\
 & =
{\rm tr} \left\{ {\bf G} \cdot {\bf R}_{R}  \cdot {\bf
G}^H \right\} = {\rm tr} \left\{ {\bf
G}^H  \cdot {\bf G} \cdot {\bf R}_{R}  \right\} \label{TrPo}
\end{align}
where $\| \cdot \|_2$ denotes the Euclidian norm of a vector and
${\bf R}_{R} = \mathbb{E} \{ {\bf r} \cdot {\bf r}^H \}$ is the
covariance matrix of ${\bf r}$ which is given by
\begin{align}
{\bf R}_{R} = {\bf h}_1^{(f)} \cdot \left( {\bf h}_1^{(f)}
\right)^H \cdot P_{T,1} + {\bf h}_2^{(f)} \cdot \left( {\bf
h}_2^{(f)} \right)^H \cdot P_{T,2} + {\bf R}_{N,R}.
\end{align}
Next, we use the equality $\rm tr(\mathbf A^{\it H} \cdot \mathbf
B) = {\rm vec}(\mathbf A)^H \cdot {\rm vec}(\mathbf B)$, which
holds for any arbitrary square matrices $\mathbf A$ and $\mathbf
B$, and where ${\rm vec} ( \cdot )$ stands for the vectorization
operation that transforms a matrix into a long vector stacking the
columns of the matrix one after another. Then, the total transmit
power of the relay  \eqref{TrPo} can be equivalently expressed as
\begin{align} \mathbb{E} \{\| {\bf{\bar{r}}} \|_2^2 \} & =
{\rm vec}(\mathbf G)^{\it H} \cdot {\rm vec}( \bf G \cdot {\bf
R}_{R}). \label{TrPo1}
\end{align}
Finally, using the equality ${\rm vec}(\mathbf A \cdot \mathbf B)
= ( \mathbf B^T \otimes \mathbf I_L) \cdot {\rm vec}(\mathbf A)$,
which is valid for any arbitrary square matrices $\mathbf A_{L
\times L}$ and $\mathbf B_{L \times L}$, and where $\otimes$
denotes the Kronecker product, \eqref{TrPo1} can be equivalently
rewritten as the following quadratic form
\begin{align}
\mathbb{E} \{\| {\bf{\bar{r}}} \|_2^2 \} & = {\bf g}^H \cdot
\underbrace{\left({\bf R}_{R}^T \otimes {\bf I}_{M_R}
\right)}_{\bf Q} \cdot {\bf g} = {\bf g}^H \cdot {\bf Q} \cdot
{\bf g}
\end{align}
where ${\bf g} = {\rm vec} \{ {\bf G} \}$.

In the second phase, the terminals receive the relay's
transmission via the (backward) channels $({\bf h}_1^{(b)})^T$ and
$({\bf h}_2^{(b)})^T$ (in the special case when reciprocity holds
we have ${\bf h}_i^{(b)} = {\bf h}_i^{(f)}$ for $i=1, 2$).
Consequently, the received signals $y_i$, $i=1, 2$ at both
terminals can be expressed, respectively, as
\begin{align}
y_1 &= h_{1,1}^{(e)} \cdot x_1 + h_{1,2}^{(e)} \cdot x_2 +
\tilde{n}_1 \\
y_2 &= h_{2,2}^{(e)} \cdot x_2 + h_{2,1}^{(e)} \cdot x_1 +
\tilde{n}_2
\end{align}
where $h_{i,j}^{(e)} = \left( {\bf h}_i^{(b)} \right)^T \cdot {\bf
G} \cdot {\bf h}_j^{(f)}$ is the effective channel between
terminal $i$ and terminal $j$ for $i,j=1,2$ and $\tilde{n}_i =
\left( {\bf h}_i^{(b)} \right)^T \cdot {\bf G} \cdot {\bf n}_{R} +
n_i$ represents the effective noise contribution at terminal $i$
which comprises the terminal's own noise as well as the forwarded
relay noise. The first term in the received signal of each
terminal represents the self-interference, which can be subtracted
by the terminal since its own transmitted signal is known. The
required channel knowledge for this step can be easily obtained,
for example, via the Least Squares (LS) compound channel estimator
described in~\cite{RH:Tence}.

After the cancellation of the self-interference, the two-way
relaying system is decoupled into two parallel single-user SISO
systems. Consequently, the rate $r_i$ of terminal $i$ can be
expressed as
\begin{align}
\label{rate}
r_i = \frac{1}{2} {\rm ld} \left( 1 + \frac{P_{R,i}}{\tilde
P_{N,i}} \right) = \frac{1}{2} {\rm ld} \left( \frac{\tilde
P_{R,i}}{\tilde P_{N,i}} \right)
\end{align}
where ${\rm ld}(\cdot)$ denotes the logarithm of base two,
$P_{R,i}$ and $\tilde P_{N,i}$ are the powers of the desired
signal and the effective noise term at terminal $i$, respectively,
and $\tilde P_{R,i} = P_{R,i} + \tilde P_{N,i}$. Specifically,
$P_{R,1} = \mathbb{E} \left\{ \left| h_{1,2}^{(e)} \cdot x_2
\right|^2 \right\}$, $P_{R,2} = \mathbb{E} \left\{ \left|
h_{2,1}^{(e)} \cdot x_1 \right|^2 \right\}$, and $\tilde{P}_{N,i}
= \mathbb{E} \left\{ | \tilde{n}_i |^2 \right\}$ for $i=1, 2$.
Note that the factor $1/2$ results from the two time slots needed
for the bidirectional transmission. The powers of the desired
signal and the effective noise term at terminal $i$ can be
equivalently expressed as
\begin{align}
P_{R,1} & =  \mathbb{E} \left\{ \left| h_{1,2}^{(e)} \cdot x_2
\right|^2 \right\} =P_{T,2} \left|
\left( {\mathbf h}_1^{(b)} \right)^T \cdot \mathbf G \cdot
\mathbf {h}_2^{(f)}\right |^2 \label{eqn_k21}  \\
P_{R,2} & = \mathbb{E} \left\{ \left| h_{2,1}^{(e)} \cdot x_1
\right|^2 \right\}  =P_{T,1} \left|
\left( {\mathbf h}_2^{(b)} \right)^T \cdot \mathbf G \cdot
\mathbf {h}_1^{(f)}\right |^2 \label{eqn_k21}  \\
\tilde{P}_{N,i} & = \mathbb{E} \left\{ \left|\left( {\bf
h}_i^{(b)} \right)^T \cdot \mathbf G \cdot \mathbf n_{R} + n_i
\right|^2 \right\} \\ &= \left( {\bf h}_i^{(b)} \right)^T \cdot
\mathbf G \cdot \mathbf R_{N,R} \cdot \mathbf G^H \left( {\bf
h}_i^{(b)} \right)^* + P_{N,i}
\end{align}
where the expectation is taken with respect to the transmit
signals and also the additional noise terms. Moreover, these
powers can be further expressed as quadratic forms in $\bf g$. For
this goal, first note that by using the following equality
\begin{equation}
{\rm vec}(\mathbf A \cdot \mathbf B \cdot \mathbf C) = (\mathbf
C^T \otimes \mathbf A) \cdot {\rm vec}(\mathbf B)
\label{Equation1}
\end{equation}
which is valid for any arbitrary matrices $\mathbf A$, $\mathbf B$
and $\mathbf C$ of compatible dimensions, the term $\left(
{\mathbf h}_i^{(b)} \right)^T \cdot \mathbf G \cdot \mathbf
{h}_j^{(f)}$ can be modified as follows
\begin{align}
\left( {\mathbf h}_i^{(b)} \right)^T \cdot \mathbf G \cdot \mathbf
{h}_j^{(f)}& = {\rm vec}\left(\left( {\mathbf h}_i^{(b)} \right)^T
\cdot \mathbf G \cdot \mathbf {h}_j^{(f)}\right) \\
& =  \left( \left( {\mathbf h}_j^{(f)} \right)^T \otimes \left(
{\mathbf h}_i^{(b)} \right)^T \right) \cdot {\rm vec} (\mathbf G).
\label{vecform}
\end{align}
Using \eqref{vecform}, the power of the desired signal at the
first terminal can be expressed as
\begin{align}
P_{R,1} & = \mathbf g^H \cdot \left( \left( {\mathbf h}_2^{(f)}
\right)^T \otimes \left( {\mathbf h}_1^{(b)} \right)^T \right)^H
\cdot \left( \left( {\mathbf h}_2^{(f)} \right)^T \otimes \left(
{\mathbf h}_1^{(b)} \right)^T \right) \cdot \mathbf g \cdot
P_{T,2} . \label{powq1}
\end{align}
Finally, applying the equality $(\mathbf A \otimes \mathbf B)
\cdot(\mathbf C \otimes \mathbf D) = (\mathbf A \cdot \mathbf C)
\otimes (\mathbf B \cdot \mathbf D)$ to \eqref{powq1} which is
valid for any arbitrary matrices $\mathbf A$, $\mathbf B$,
$\mathbf C$ and $\mathbf D$ of agreed dimensions, $P_{R,1}$ can be
expressed as the following quadratic form
\begin{align}
P_{R,1} & = {\bf g}^H \cdot \left[ \left( {\bf h}_2^{(f)} \cdot
\left( {\bf h}_2^{(f)} \right)^H \right) \otimes \left( {\bf
h}_1^{(b)} \cdot \left( {\bf h}_1^{(b)} \right)^H \right)
\right]^T \mathbf g \cdot P_{T,2} .
\end{align}
Similarly, $P_{R,2}$ can be obtained.

By defining the matrices ${\bf K}_{2,1}$,  ${\bf K}_{1,2}$ as
follows
\begin{align}
{\bf K}_{2,1} & = \left[ \left( {\bf h}_2^{(f)} \cdot \left( {\bf
h}_2^{(f)} \right)^H \right) \otimes \left( {\bf h}_1^{(b)} \cdot
\left( {\bf h}_1^{(b)} \right)^H \right) \right]^T \\
{\bf K}_{1,2} & = \left[ \left( {\bf h}_1^{(f)} \cdot \left( {\bf
h}_1^{(f)} \right)^H \right) \otimes \left( {\bf h}_2^{(b)} \cdot
\left( {\bf h}_2^{(b)} \right)^H \right) \right]^T
\end{align}
the powers of the desired signal can be expressed as
\begin{align}
P_{R,1} & = {\bf g}^H \cdot {\bf K}_{2,1} \cdot {\bf g} \cdot
P_{T,2} \label{eqn_k21}  \\
P_{R,2} & = {\bf g}^H \cdot {\bf K}_{1,2} \cdot {\bf g} \cdot
P_{T,1}. \label{eqn_k12}
\end{align}
As the last step, the effective noise $\tilde{P}_{N,i}$ can be
converted into a quadratic form through the following train of
equalities
\begin{align}
\tilde{P}_{N,i} & = \mathbb{E} \left\{ \left|\left( {\bf
h}_i^{(b)} \right)^T \cdot \mathbf G \cdot \mathbf n_{R} + n_i
\right|^2 \right\}   \nonumber \\ &= \left( {\bf h}_i^{(b)}
\right)^T \cdot \mathbf G \cdot \mathbf R_{N,R} \cdot \mathbf G^H
\left( {\bf h}_i^{(b)} \right)^* + P_{N,i}
\nonumber \\
&= {\rm tr}\left(\mathbf G^H  \cdot \left( {\bf h}_i^{(b)}
\right)^* \cdot \left( {\bf h}_i^{(b)} \right)^T \cdot \mathbf G
\cdot \mathbf R_{N,R} \right) + P_{N,i}
\label{l3} \\
&= {\rm vec}(\mathbf G)^H \cdot {\rm vec}\left(\left( {\bf
h}_i^{(b)} \right)^* \cdot \left( {\bf h}_i^{(b)} \right)^T \cdot
\mathbf G \cdot \mathbf R_{N,R} \right)
+ P_{N,i} \label{l4} \\
&= {\rm vec}(\mathbf G)^H \cdot  \left(\mathbf R_{N,R} \otimes
\left( {\bf h}_i^{(b)} \right) \left( {\bf h}_i^{(b)} \right)^H
\right)^T \cdot {\rm vec}  (\mathbf G) + P_{N,i}
\label{l5} \\
&= \mathbf g^H \cdot \mathbf J_i \cdot \mathbf g + P_{N,i}
\label{eqn_pnti}
\end{align}
where \eqref{l4} is obtained from \eqref{l3} by applying the the
equality ${\rm tr}(\mathbf A^H \cdot \mathbf B)= {\rm vec}
(\mathbf A)^H \cdot {\rm vec}(\mathbf B)$, which is valid for any
arbitrary square matrices matrices $\mathbf A_{L \times L}$ and
$\mathbf B_{L \times L}$,  equation \eqref{l5} is obtained from
\eqref{l4} by applying the equality \eqref{Equation1}, and the
matrix ${\bf J}_i$ is defined as
\begin{align}
{\bf J}_i & = \left[ {\bf R}_{N,R} \otimes \left( {\bf h}_i^{(b)}
\cdot \left( {\bf h}_i^{(b)} \right)^H \right) \right]^T.
\label{eqn_def_ji}
\end{align}

\section{Problem statement}
Our goal is to find the relay amplification matrix ${\bf G}$ which
maximizes the sum-rate $r_1 + r_2$ subject to a power constraint
at the relay. For convenience we express the objective function
and its solution in terms of ${\bf g} = {\rm vec} \{ {\bf G} \}$.
Then the power constrained sum-rate maximization problem can be
expressed as
\begin{align}
{\bf g}_{\rm opt} = \mathop{\arg \max}_{\bf g} \Big( r_1 + r_2
\Big) \; \mbox{subject to} \; {\bf g}^H \cdot {\bf Q} \cdot {\bf
g} \leq P_{T,R}
\end{align}
where $P_{T,R}$ is the allowed transmit power at the relay. Using
the definitions from the previous section, this optimization
problem can be rewritten as
\begin{align}
{\bf g}_{\rm opt} & = \mathop{\arg \max}_{{\bf g} | {\bf g}^H
\cdot {\bf Q} \cdot {\bf g} \leq P_{T,R} } \frac{1}{2} {\rm ld}
\left[ \left( 1 + \frac{P_{R,1}}{\tilde P_{N,1}} \right) \cdot
\left( 1 + \frac{P_{R,2}}{\tilde P_{N,2}} \right) \right]
\nonumber \\
& = \mathop{\arg \max}_{{\bf g} | {\bf g}^H \cdot {\bf Q} \cdot
{\bf g} \leq P_{T,R} } \left( 1 + \frac{P_{R,1}}{\tilde P_{N,1}}
\right) \cdot \left( 1 + \frac{P_{R,2}}{\tilde P_{N,2}} \right)
\label{eqn_optp_i1} \\
& = \mathop{\arg \max}_{{\bf g} | {\bf g}^H \cdot {\bf Q} \cdot
{\bf g} \leq P_{T,R} } \frac{\tilde P_{R,1}}{\tilde P_{N,1}} \cdot
\frac{\tilde P_{R,2}}{\tilde P_{N,2}} \label{eqn_optp_i2}
\end{align}
where we have used the fact that $0.5 \cdot {\rm ld} (x)$ is a
monotonic function in $x \in \mathbb{R}^+$ and $\tilde
P_{R,i},i=1,2$ is defined after \eqref{rate}.

It is worth noting that the inequality constraint in this
optimization problem has to be active at the optimal point. This
can be easily shown by contradiction. Assume ${\bf g}_{\rm opt}$
satisfies ${\bf g}_{\rm opt}^H \cdot {\bf Q} \cdot {\bf g}_{\rm
opt} < P_{T,R}$. Then we can find a constant $c > 1$ such that
${\bar{\bf g}}_{\rm opt} = c \cdot {\bf g}_{\rm opt}$ satisfies
${\bar{\bf g}}_{\rm opt}^H \cdot {\bf Q} \cdot {\bar{\bf g}}_{\rm
opt} = P_{T,R}$. However, inserting ${\bar{\bf g}}_{\rm opt}$ in
the objective function of~\eqref{eqn_optp_i1}, we obtain
\begin{align}
\left( 1 + \frac{c^2 \cdot {\bf g}_{\rm opt}^H {\bf K}_{2,1} {\bf
g}_{\rm opt} P_{T,2}}{c^2 \cdot {\bf g}_{\rm opt}^H {\bf J}_1 {\bf
g}_{\rm opt} + P_{N,1}} \right) \cdot \left( 1 + \frac{c^2 \cdot
{\bf g}_{\rm opt}^H {\bf K}_{1,2} {\bf g}_{\rm opt}  P_{T,1}}{c^2
\cdot {\bf g}_{\rm opt}^H {\bf
J}_2 {\bf g}_{\rm opt} + P_{N,2}} \right) \nonumber \\
= \left( 1 + \frac{{\bf g}_{\rm opt}^H {\bf K}_{2,1} {\bf g}_{\rm
opt} P_{T,2}}{{\bf g}_{\rm opt}^H {\bf J}_1 {\bf g}_{\rm opt} +
\frac{P_{N,1}} {c^2}} \right) \cdot \left( 1 + \frac{{\bf g}_{\rm
opt}^H {\bf K}_{1,2} {\bf g}_{\rm opt} P_{T,1}}{{\bf g}_{\rm
opt}^H {\bf J}_2 {\bf g}_{\rm opt} + \frac{P_{N,2}}{c^2}} \right)
\end{align}
which is monotonically increasing in $c$. Since we have $c > 1$,
the vector ${\bar{\bf g}}_{\rm opt}$ provides a larger value of
the objective functions than ${\bf g}_{\rm opt}$ which contradicts
the assumption that ${\bf g}_{\rm opt}$ was optimal.

As a result, we have shown that the optimal vector ${\bf g}_{\rm
opt}$ must satisfy the total power constraint of the problem with
equality, i.e., ${\bf g}_{\rm opt}^H \cdot {\bf Q} \cdot {\bf
g}_{\rm opt} = P_{T,R}$. Using this fact, the inequality
constraint in the problem \eqref{eqn_optp_i2}  can be replaced by
the constraint ${\bf g}^H \cdot{\bf Q} \cdot {\bf g}= P_{T,R}$.
This enables us to substitute the constant term $P_{N,i}$, which
appears in the effective noise power at terminal $i$
\eqref{eqn_pnti}, with the quadratic term of ${\bf g}_{\rm opt}^H
\cdot {\bf Q} \cdot {\bf g}_{\rm opt} \cdot (P_{N,i}/ P_{T,R})$.
This leads to an equivalent homogeneous expression for the ratio
of ${\tilde P_{R,1}} / {\tilde P_{N,1}}, i=1,2$. Thus,
by using such substitution, $\tilde P_{N,i}$, $i=1,2$
from~\eqref{eqn_pnti} can be equivalently written as
\begin{align}
\tilde P_{N,i} = {\bf g}^H \cdot {\bf B}_i \cdot {\bf g}, \quad
i=1,2
\end{align}
where ${\bf B}_i$ is given by
\begin{align}
{\bf B}_i = {\bf J}_i + \frac{P_{N,i}}{P_{T,R}} \cdot {\bf Q} .
\label{eqn_jitilde}
\end{align}
Inserting~\eqref{eqn_k21}, \eqref{eqn_k12}, and
\eqref{eqn_jitilde} into~\eqref{eqn_optp_i2}, the optimization
problem becomes
\begin{align}
{\bf g}_{\rm opt} & = \mathop{\arg \max}_{{\bf g} | {\bf g}^H
\cdot {\bf Q} \cdot {\bf g} = P_{T,R}} \frac{{\bf g}^H \cdot {\bf
A}_{1} \cdot {\bf g}}{{\bf g}^H \cdot {\bf B}_1 \cdot {\bf g}}
\cdot \frac{{\bf g}^H \cdot {\bf A}_{2} \cdot {\bf g}}{{\bf g}^H
\cdot {\bf B}_2 \cdot {\bf g}} \label{eqn_costf_i3}
\end{align}
where we have defined the new matrices ${\bf A}_{1} = {\bf
K}_{2,1} \cdot P_{T,2} + {\bf B}_1$ and ${\bf A}_{2} = {\bf
K}_{1,2} \cdot P_{T,1} + {\bf B}_2$.

As a final simplifying step we observe that the objective function
of~\eqref{eqn_costf_i3} is homogeneous in ${\bf g}$, meaning that
an arbitrary rescaling of ${\bf g}$ has no effect on the value of
the objective functions. Consequently, the equality constraint can
be dropped completely as any solution to the unconstrained problem
can be rescaled to meet the equality constraint without any loss
in terms of the objective functions. Therefore, the final form of
our problem statement is given by
\begin{align}
{\bf g}_{\rm opt} & = \mathop{\arg \max}_{\bf g} \frac{{\bf g}^H
\cdot {\bf A}_1 \cdot {\bf g}}{{\bf g}^H \cdot {\bf B}_1 \cdot
{\bf g}} \cdot \frac{{\bf g}^H \cdot {\bf A}_2 \cdot {\bf g}}{{\bf
g}^H \cdot {\bf B}_2 \cdot {\bf g}} . \label{eqn_costf_final}
\end{align}

Note that from their definitions it is obvious that ${\bf A}_i$,
$i=1,2$ and ${\bf B}_i$, $i=1,2$ are positive definite matrices.
Therefore, the optimization problem~\eqref{eqn_costf_final} can be
interpreted as the product of two Rayleigh quotients. Moreover, it
can be expressed as a DC programming problem. Indeed, as we will
show later in details, by expressing the
problem~\eqref{eqn_costf_final} as a rank constrained problem and
then dropping the rank constraint and also taking the logarithm of
the objective function, the objective function of the resulting
problem can be written as the summation of two concave functions
with positive signs and two concave functions with negative signs.
Thus, the objective of the equivalent problem is, in fact, the
difference of convex functions which is in general non-convex, and
the available algorithms in the literature for solving such DC
programming problems are based on the so-called branch-and-bound
method that does not have any polynomial time computational
complexity guarantees \cite{JL10}-\cite{Tuy2}. However, as we show
next, the problem~\eqref{eqn_costf_final} can be parameterized in
such a way that there exist simple polynomial time solutions.

\section{Polynomial-Time Solution for the Sum-Rate Maximization
Problem in Two-Way AF MIMO Relaying} Since the problem
\eqref{eqn_costf_final} is homogenous, without loss of generality,
we can fix the quadratic term ${\bf g}^H \cdot {\bf B}_1 \cdot
{\bf g}$ to be equal to one at the optimal point. By doing so and
also by defining the additional variables $\tau$ and $\beta$,
the problem \eqref{eqn_costf_final} can be equivalently recast as
\begin{eqnarray}
\max\limits_{{\bf g}, \tau, \beta} \!\!\!&&\!\!\! {\bf g}^H \cdot
{\bf A}_1 \cdot {\bf g} \cdot \frac{\tau}{\beta}
\nonumber \\
\!\!\!&&\!\!\! {\bf g}^H \cdot {\bf B}_1 \cdot {\bf g} = 1
\nonumber \\
\!\!\!&&\!\!\! {\bf g}^H \cdot {\bf A}_2 \cdot {\bf g} = \tau
\nonumber \\
\!\!\!&&\!\!\! {\bf g}^H \cdot {\bf B}_2 \cdot {\bf g} = \beta
\label{main_problem20}
\end{eqnarray}

Using the fact that the quadratic function ${\bf g}^H \cdot {\bf
B}_1 \cdot {\bf g}$ is set to one, one can easily check that the
problem~\eqref{main_problem20} is feasible if and only if $\tau
\in [\lambda_{\min}(\mathbf B_1^{-1} \mathbf A_2), \lambda_{\max}
(\mathbf B_1^{-1} \mathbf A_2)]$ and $\beta \in [\lambda_{\min}
(\mathbf B_1^{-1} \mathbf B_2), \lambda_{\max}(\mathbf B_1^{-1}
\mathbf B_2)]$ where $\lambda_{\min} (\cdot)$ and $\lambda_{\max}
(\cdot)$ denote the smallest and the largest eigenvalues operator,
respectively. By introducing the matrix ${\bf X} \triangleq {\bf
g} \cdot {\bf g}^H$ and observing that for any arbitrary matrix
${\bf Y}$, the equation ${\bf g}^H \cdot {\bf Y} \cdot {\bf g} =
{\rm tr} ({\bf Y} \cdot {\bf g} \cdot {\bf g}^H)$ holds, the
optimization problem \eqref{main_problem20} can be equivalently
expressed as
\begin{eqnarray}
\max\limits_{\bf X, \tau,\beta} \!\!\!&&\!\!\!  {\rm tr} ({\bf
A}_1 \cdot {\bf X}) \cdot \frac{\tau}{\beta} \nonumber \\
\!\!\!&&\!\!\! {\rm tr} ({\bf B}_1 \cdot {\bf X}) = 1 \nonumber \\
\!\!\!&&\!\!\! {\rm tr} ({\bf A}_2 \cdot {\bf X}) = \tau
\nonumber \\
\!\!\!&&\!\!\! {\rm tr} ({\bf B}_2 \cdot {\bf X}) = \beta
\nonumber \\
\!\!\!&&\!\!\! {\rm rank} ({\bf X}) = 1, \ \ \ {\bf X} \succeq
{\bf 0} . \label{main_problem3}
\end{eqnarray}

In what follows, we explain the possibility of dropping the
rank-one constraint in the problem \eqref{main_problem3} and then
extracting the exact solution for the original problem
\eqref{main_problem3} based on the solution of the rank relaxed
problem. To this end, let ${\bf X}_{\tau, \beta}$ denote the
optimal solution of the optimization problem \eqref{main_problem3}
with respect to $\bf X$ for fixed values of $\tau$ and $\beta$ and
without considering the rank-one constraint. It is known that the
strong duality for a QCQP problem with three or less constraints
is satisfied \cite{Daniel}. Based on this fact, the strong duality
holds for the problem \eqref{main_problem20}, which for fixed
variables $\tau$ and $\beta$ is equivalent to QCQP with three
constraints. Since the problem \eqref{main_problem3} is equivalent
to the problem \eqref{main_problem20}, the strong duality also
holds for \eqref{main_problem3} for fixed $\tau$ and $\beta$. As a
result, a rank-one solution of the problem \eqref{main_problem3}
can always be constructed based on ${\bf X}_{\tau, \beta}$ for
fixed $\tau$ and $\beta$. Thus, for fixed $\tau$ and $\beta$, the
optimal value of the problem \eqref{main_problem3} with respect to
$\bf X$ is independent of the rank-one constant. It enables us to
drop the rank-one constraint in the problem \eqref{main_problem3},
solve the relaxed problem, and then construct an optimal rank-one
solution once the optimal ${\bf X}_{\rm opt}$, $\tau_{\rm opt}$,
and $\beta_{\rm opt}$ are obtained. Dropping the rank-one
constraint results in the following optimization problem
\begin{eqnarray}
\max\limits_{\mathbf X,\tau,\beta} \!\!\!&&\!\!\! {\rm tr} ({\bf
A}_1 \cdot {\bf X}) \cdot \frac{\tau}{\beta} \nonumber \\
\!\!\!&&\!\!\! {\rm tr} ({\bf B}_1 \cdot {\bf X}) = 1  \nonumber \\
\!\!\!&&\!\!\! {\rm tr} ({\bf A}_2 \cdot {\bf X}) = \tau \nonumber \\
\!\!\!&&\!\!\! {\rm tr} ({\bf B}_2 \cdot {\bf X}) = \beta \nonumber \\
\!\!\!&&\!\!\! {\bf X} \succeq {\bf 0} . \label{main_problem4}
\end{eqnarray}

Due to the fact that the matrix $\mathbf A_1$ is positive definite
and $\bf X$ is positive semi-definite, the function ${\rm tr}
({\bf A}_1 \cdot {\bf X})$ is always positive. The latter happens
since the matrix $\mathbf X$ cannot be equal to a zero matrix due
to the constraint ${\rm tr} ({\bf B}_1 \cdot {\bf X}) = 1$.
Moreover, since the values $\lambda_{\min}(\mathbf B_1^{-1}
\mathbf A_2)$ and $\lambda_{\min}(\mathbf B_1^{-1} \mathbf B_2)$
are necessarily positive, the variables $\tau$ and $\beta$ are
also positive. The task of maximizing the objective function in
the problem~\eqref{main_problem4} is equivalent to maximizing the
logarithm of this objective function because $\log(x)$ is a
strictly increasing function and the objective function in
\eqref{main_problem4} is positive. Therefore, the optimization
problem~\eqref{main_problem4} can be equivalently rewritten as
\begin{eqnarray}
\max\limits_{\bf X, \tau, \beta} \!\!\!&&\!\!\! \log ({\rm tr}
({\bf A}_1 \cdot {\bf X})) + \log (\tau) - \log (\beta) \nonumber \\
\!\!\!&&\!\!\! {\rm tr} ({\bf B}_1 \cdot {\bf X} ) = 1 \nonumber \\
\!\!\!&&\!\!\! {\rm tr} ({\bf A}_2 \cdot {\bf X} ) = \tau \nonumber \\
\!\!\!&&\!\!\! {\rm tr} ({\bf B}_2 \cdot {\bf X} ) = \beta \nonumber \\
\!\!\!&&\!\!\! {\bf X} \succeq {\bf 0} \label{final_relaxed0}
\end{eqnarray}
%Since $\log ({\rm tr} ({\bf A}_1 \cdot {\bf X}))$, $\log (\tau)$
%and $\log (\beta)$ are all concave functions with respect to
%$\mathbf X$, $\tau$ and $\beta$, the problem
%\eqref{final_relaxed0} can be expressed as the minimization of the
%difference of two convex functions.
Note that dropping the rank-one constraint enabled us to write our
optimization problem as a DC programming problem, where the fact
that $\log ({\rm tr} ({\bf A}_1 \cdot {\bf X}))$ in the objective
of \eqref{final_relaxed0} is a concave function is also
considered. Although the problem \eqref{final_relaxed0} boils down
to the known family of DC programming problems, still there exists
no solution for such DC programming problems with guaranteed
polynomial time complexity. However, the problem
\eqref{final_relaxed0} has a very particular structure, such as,
all the constraints are convex and the terms $\log ({\rm tr} ({\bf
A}_1 \cdot {\bf X}))$ and $\log (\tau)$ in the objective are
concave. Thus, the only term that makes the problem overall
non-convex is the term $-\log (\beta)$ in the objective. If $-\log
(\beta)$ is piece-wise linearized over a finite number of
intervals\footnote{As explained before, the parameter $\beta$ can
take values only in a finite interval. Thus, a finite number of
linearization intervals for $-\log (\beta)$ is needed.}, then the
objective function becomes concave on these intervals and the
whole problem \eqref{final_relaxed0} becomes convex. The resulting
convex problems over different linearization intervals for $-\log
(\beta)$ can be solved efficiently in polynomial time, and then,
the suboptimal solution of the problem \eqref{final_relaxed0} can
be found. The fact that such a solution is suboptimal follows from
the linearization, which has a finite accuracy. The smaller the
intervals are, the more accurate becomes the solution of
\eqref{final_relaxed0}. This solution is also not the most
efficient in terms of complexity. Thus, we develop another method
(the POTDC algorithm) which makes it possible to solve the
problem~\eqref{final_relaxed0} in a more efficient way.

To fulfil this goal, we introduce a new additional variable $t$,
which makes it possible to express the problem
\eqref{final_relaxed0} equivalently as
\begin{eqnarray}
\max\limits_{\bf X, \tau, \beta} \!\!\!&&\!\!\! \log({\rm tr}
({\bf A}_1 \cdot {\bf X})) + \log( \tau) - t \nonumber \\
\!\!\!&&\!\!\! {\rm tr} ({\bf B}_1 \cdot {\bf X} ) = 1
\nonumber \\
\!\!\!&&\!\!\! {\rm tr} ({\bf A}_2 \cdot {\bf X} ) = \tau
\nonumber \\
\!\!\!&&\!\!\! {\rm tr} ({\bf B}_2 \cdot {\bf X} ) = \beta
\nonumber \\
\!\!\!&&\!\!\! \log (\beta) \leq t \nonumber \\
\!\!\!&&\!\!\! {\bf X} \succeq {\bf 0} . \label{final_relaxed}
\end{eqnarray}
The objective function of the optimization problem
\eqref{final_relaxed} is concave and all the constraints except
the constraint $\log(\beta) \leq t$ are convex. Thus, we can
develop an iterative method that is different to the
aforementioned piece-wise linearization-based method, and is based
on linearizing the non-convex term $\log (\beta)$ in the
constraint $\log(\beta) \leq t$ around a suitably selected point
in each iteration. More specifically, the linearizing point in
each iteration is selected such that the iterative algorithm gets
closer to optimal point in every iteration. Roughly speaking, the
main idea of this iterative method is similar to the gradient
based methods. In the first iteration, we start with an arbitrary
point selected in the interval $[\lambda_{\min}(\mathbf B_1^{-1}
\mathbf B_2), \lambda_{\max}(\mathbf B_1^{-1} \mathbf B_2)]$ and
denoted as $\beta_c$. Then the non-convex function $\log(\beta)$
can be replaced by its linear approximation around this point
$\beta_c$, that is,
\begin{equation} \label{lin}
\log (\beta) \approx \log(\beta_c) + \frac{1}{\beta_c} (\beta -
\beta_c)
\end{equation}
which results in the following convex optimization problem
\begin{eqnarray}
\max\limits_{\bf X, \tau, \beta} \!\!\!&&\!\!\! \log({\rm tr}
({\bf A}_1 \cdot {\bf X})) + \log( \tau) - t \nonumber \\
\!\!\!&&\!\!\! {\rm tr} ({\bf B}_1 \cdot {\bf X} ) = 1
\nonumber \\
\!\!\!&&\!\!\! {\rm tr} ({\bf A}_2 \cdot {\bf X} ) = \tau \
\nonumber \\
\!\!\!&&\!\!\! {\rm tr} ({\bf B}_2 \cdot {\bf X} ) = \beta
\nonumber \\
\!\!\!&&\!\!\! \log(\beta_c) + \frac{1}{\beta_c}(\beta - \beta_c)
\leq t \nonumber \\
\!\!\!&&\!\!\! {\bf X} \succeq {\bf 0} . \label{linearized}
\end{eqnarray}

The problem \eqref{linearized} can be efficiently solved by means
of the interior-point based numerical methods. Once the optimal
solution of this problem in the first iteration, denoted as
$\mathbf X_{\rm opt}^{(1)}$, $\tau_{\rm opt}^{(1)}$ and
$\beta_{\rm opt}^{(1)}$, is found, the algorithm proceeds to the
second iteration by replacing the function $\log (\beta)$ by its
linear approximation around $\beta_{\rm opt}^{(1)}$ found from the
previous (first) iteration. Fig.~\ref{linearization} shows how
$\log(\beta)$ is replaced by its linear approximation around
$\beta_c$ where $\beta_{\rm opt}$ is the optimal value of $\beta$
obtained through solving \eqref{linearized} using such a linear
approximation. In the second iteration, the resulting optimization
problem has the same structure as the problem \eqref{linearized}
in which $\beta_c$ has to be set to $\beta_{\rm opt}^{(1)}$
obtained from the first iteration. This process continues and
every iteration is obtained by replacing $\log (\beta)$ at the
iteration $k$ by its linearization of type \eqref{lin} around
$\beta_{\rm opt}^{(k-1)}$ found from the iteration $k-1$. The
POTDC algorithm for solving the problem \eqref{final_relaxed} is
summarized in Algorithm~1.

\begin{algorithm}
\caption{The POTDC algorithm for solving the optimization problem
\eqref{final_relaxed}} \label{pa_pa_al}
\begin{tabular}{l*{6}{c}}
\hline
\end{tabular}
\begin{algorithmic}
\STATE \textbf{Initialize:} Select an arbitrary $\beta_c$ from
theinterval $ [\lambda_{\min} ({\bf B}_1^{-1} {\bf B}_2 ),
\lambda_{\max} ({\bf B}_1^{-1} {\bf B}_2 ) ]$, set the counter $k$
to be equal to $1$ and choose an accuracy parameter $\epsilon$.

\WHILE{The difference between the values of the objective function
in two consecutive iterations is larger than $\epsilon$.}

\STATE Use the linearization of type~\eqref{lin} and solve the
following optimization problem
\begin{eqnarray}
\max\limits_{\bf X, \tau, \beta} \!\!\!&&\!\!\! \log({\rm tr}
({\bf A}_1 \cdot {\bf X})) + \log( \tau) - t \nonumber \\
\!\!\!&&\!\!\! {\rm tr} ({\bf B}_1 \cdot {\bf X} ) = 1
\nonumber \\
\!\!\!&&\!\!\! {\rm tr} ({\bf A}_2 \cdot {\bf X} ) = \tau \
\nonumber \\
\!\!\!&&\!\!\! {\rm tr} ({\bf B}_2 \cdot {\bf X} ) = \beta
\nonumber \\
\!\!\!&&\!\!\! \log(\beta_c) + \frac{1}{\beta_c}(\beta - \beta_c)
\leq t \nonumber \\
\!\!\!&&\!\!\! {\bf X} \succeq {\bf 0} . \label{linearizedAlg}
\end{eqnarray}
to obtain ${\bf X}_{\rm opt}^{(k)}$, $\tau_{\rm opt}^{(k)}$, and
$\beta_{\rm opt}^{(k)}$.

\STATE $k = k + 1$

\STATE Set ${\bf X}_{\rm opt} := {\bf X}_{\rm opt}^{(k)}$, and
$\beta_{c}:=\beta_{\rm opt}^{(k)}$.

\ENDWHILE

\STATE \textbf{Output:} ${\bf X}_{\rm opt}$.
\end{algorithmic}
\end{algorithm}

The following two lemmas regarding the proposed POTDC algorithm
are of interest. First, the termination condition in the POTDC
algorithm is guaranteed to be satisfied due to the following lemma
which states that by choosing $\beta_c$ in the above proposed
manner, the optimal values of the objective function
of~\eqref{linearized} for ${\bf X}_{\rm opt}^{(k)}$, $\tau_{\rm
opt}^{(k)}$, and $\beta_{\rm opt}^{(k)}$ are non-decreasing.

\textbf{Lemma~1:} The optimal values of the objective function of
the optimization problem~\eqref{linearized} obtained over the
iterations of the POTDC algorithm are non-decreasing.

\begin{proof}
Considering the linearized problem \eqref{linearized} in the
iteration $k+1$, it is easy to verify that ${\bf X}_{\rm
opt}^{(k)}$, $\tau_{\rm opt}^{(k)}$, and $\beta_{\rm opt}^{(k)}$
give a feasible point for this problem. Therefore, it can be
concluded that  the optimal value at the iteration $k+1$ must be
greater than or equal to the optimal value in the iteration $k$
which completes the proof.
\end{proof}

Second, it is guaranteed that the solution obtained using the
POTDC algorithm is optimal due to the following lemma.

\textbf{Lemma~2:} The solution obtained using the POTDC algorithm
satisfies the Karush–-Kuhn–-Tucker (KKT) conditions.

\begin{proof}
This lemma follows straightforwardly from a similar proposition in
\cite{Beck}.
\end{proof}

As soon as the solution of the relaxed
problem~\eqref{final_relaxed} is found, the solution of the
original problem~\eqref{main_problem20}, which is equivalent to
the solution of the sum-rate maximization
problem~\eqref{eqn_costf_final}, can be found using one of the
existing methods for extracting a rank one solution. Among the
existing methods are the ones based on solving the dual problem
\cite{BeckEldar}, which exploits the fact that the original
problem \eqref{main_problem20} with only two constraints is
strictly feasible and has zero duality gap; the algebraic
technique of \cite{ArashSergiy}; and the rank reduction-based
technique of \cite{Daniel} which is also applicable for the
problems with three constrains. Although the solution of
\eqref{final_relaxed} is guaranteed to be optimal, it is still
left to show that this solution is also globally optimal.

\section{An Upper-Bound for the Optimal Value}
Through extensive simulations we have observed that regardless of
the initial value chosen for $\beta_c$ in the first iteration of
the POTDC algorithm, the proposed iterative method always
converges to the global optimum of the problem
\eqref{final_relaxed}. However, since the original problem is not
convex, this can not be easily verified analytically. A comparison
between the optimal value obtained by using the proposed iterative
method and also the global optimal value can be, however, done by
developing a tight upper-bound for the optimal value of the
problem and comparing the solution to such an upper-bound. Thus,
in this section, we find such an upper-bound for the optimal value
of the optimization problem \eqref{main_problem20}. For this goal,
we first consider the following lemma which gives an upper-bound
for the optimal value of the variable $\beta$ in the problem
\eqref{final_relaxed0}. This lemma will further be used for
obtaining the desired upper-bound for our problem.

\textbf{Lemma~3:} The optimal value of the variable $\beta$ in
\eqref{final_relaxed}, denoted as $\beta_{\rm opt}$ is
upper-bounded by $e^{(q^\star - p^\star)}$, where $p^\star$ is the
value of the objective function in the
problem~\eqref{final_relaxed} corresponding to any arbitrary
feasible point and $q^\star$ is the solution of the following
convex optimization problem\footnote{Note that this optimization
problem can be solved efficiently using numerical methods, for
example, interior point methods.}
\begin{eqnarray}
q^\star= \mathop{\max}\limits_{\bf X, \tau, \beta} \!\!\!&&\!\!\!
\log( {\rm tr} ( {\bf A}_1 \cdot {\bf X} )) + \log( \tau)
\nonumber \\
\!\!\!&&\!\!\! {\rm tr} ({\bf B}_1 \cdot {\bf X} ) \nonumber = 1 \\
\!\!\!&&\!\!\! {\rm tr} ({\bf A}_2 \cdot {\bf X} ) = \tau
\nonumber \\
\!\!\!&&\!\!\! {\bf X} \succeq {\bf 0} .
\label{final_relaxed_double}
\end{eqnarray}

\begin{proof}
First note that since $p^\star$ is the value of the objective
function in the problem~\eqref{final_relaxed} corresponding to an
arbitrary feasible point, it must be less than or equal to the
optimal value of problem~\eqref{final_relaxed}. By fixing the
variable $\beta$ to $\beta_{\rm opt}$ in the optimization problem
\eqref{final_relaxed}, the optimal value of the objective function
does not change. Moreover, in the aforementioned case when $\beta$
has been fixed to $\beta_{\rm opt}$, dropping the constraint ${\rm
tr} ({\bf B}_2 \cdot {\bf X} ) = \beta_{\rm opt}$ in that problem
leads to the following optimization problem
\begin{eqnarray}
\max\limits_{\bf X, \tau, \beta} \!\!\!&&\!\!\! \log ({\rm tr}
({\bf A}_1 \cdot {\bf X} )) + \log( \tau) - \log(\beta_{\rm opt})
\nonumber \\
\!\!\!&&\!\!\! {\rm tr} ({\bf B}_1 \cdot {\bf X} ) =1 \nonumber \\
\!\!\!&&\!\!\! {\rm tr} ({\bf A}_2 \cdot {\bf X} ) = \tau
\nonumber \\
\!\!\!&&\!\!\! {\bf X} \succeq {\bf 0} .
\label{final_relaxed_double0}
\end{eqnarray}
Noticing that the feasible set of the optimization problem
\eqref{final_relaxed} is a subset of the feasible set of the newly
introduced optimization problem \eqref{final_relaxed_double0}, it
is straightforward to conclude that the optimal value of the
problem \eqref{final_relaxed_double0} is bigger than or equal to
the optimal value of the problem \eqref{final_relaxed} and as a
result it is greater than or equal to $p^\star$. Using
\eqref{final_relaxed_double}, the optimal value of the
optimization problem \eqref{final_relaxed_double0} can be
expressed as $q^\star - \log(\beta_{\rm opt})$ which is bigger
than or equal to $p^\star$ and, therefore, $\beta_{\rm opt} \leq
e^{(q^\star - p^\star)}$ which completes the proof.
\end{proof}

Note that as mentioned earlier, $p^\star$ is the objective value
of the problem~\eqref{final_relaxed} that corresponds to an
arbitrary feasible point. In order to obtain the tightest possible
upper-bound for $\beta_{\rm opt}$, we choose $p^\star$ to be the
largest possible value that we already know. A suitable choice for
$p^\star$ is then the one which is obtained using the POTDC
algorithm. In other words, we choose $p^\star$ as the
corresponding objective value of the problem~\eqref{final_relaxed}
at the optimal point which is resulted from the POTDC algorithm.
Thus, we have obtained an upper-bound for $\beta_{\rm opt}$ which
makes it further possible to develop an upper-bound for the
optimal value of the optimization problem~\eqref{final_relaxed}.
To this end, we consider the only non-convex constraint of this
problem, i.e., $\log(\beta) \leq t$. Fig.~\ref{Fig1} illustrates a
subset of the feasible region corresponding to the non-convex
constraint $\log(\beta) \leq t$ where $\beta_{\min}$ equals
$\lambda_{\min}(\mathbf B_1^{-1} \mathbf B_2)$, i.e., the smallest
value of $\beta$ for which the problem \eqref{final_relaxed} is
feasible, and $\beta_{\rm max}$ is the upper-bound for the optimal
value $\beta_{\rm opt}$ given by Lemma~3. For obtaining an
upper-bound for the optimal value of the
problem~\eqref{final_relaxed}, we divide the interval $[\beta_{\rm
min}, \beta_{\rm max}]$ into $N$ sections as it is shown in
Fig.~\ref{Fig1}. Then, each section is considered separately. In
each such section, the corresponding non-convex feasible set is
replaced by its convex-hull and each corresponding optimization
problem is solved separately as well. The maximum optimal value of
such $N$ convex optimization problems is then the upper-bound.
Indeed, solving the resulting $N$ convex optimization problems and
choosing the maximum optimum value among them is equivalent to
replacing the constraint $\log(\beta) \leq t$ with the feasible
set which is described by the region above the thin line in
Fig.~\ref{Fig1}. The upper-bound becomes more and more accurate
when the number of the intervals, i.e., $N$ increases.

\section{Semi-Algebraic Solution via Generalized Eigenvectors (RAGES)}
In this section we present RAGES as an alternative solution to the
sum-rate maximization problem~\eqref{eqn_costf_final} which is
based on generalized eigenvectors. It requires a different
parameterization than the one used in the POTDC algorithm and in
some cases it is more efficient.

\subsection{Basic Approach: Generalized Eigenvectors}
To derive the link between~\eqref{eqn_costf_final} and generalized
eigenvectors we start with the necessary condition for optimality
that the gradient of~\eqref{eqn_costf_final} vanishes. Therefore,
if we find all vectors ${\bf g}$ for which the gradient of the
objective functions is zero, the global optimum must be one of
them. By using the product rule and the chain rule of
differentiation, the condition of zero gradient can be expressed
as \cite{Florian}
\begin{align}
& \frac{\tilde P_{R,2}}{\tilde P_{N,1} \cdot \tilde P_{N,2}} \cdot
{\bf A}_{1} \cdot {\bf g} + \frac{\tilde P_{R,1}}{\tilde P_{N,1}
\cdot \tilde P_{N,2}} \cdot {\bf A}_{2} \cdot {\bf g} \nonumber \\
& = \frac{\tilde P_{R,1} \cdot \tilde P_{R,2}}{\tilde P_{N,1}^2
\cdot \tilde P_{N,2}} \cdot {\bf B}_{1} \cdot {\bf g} +
\frac{\tilde P_{R,1} \cdot \tilde P_{R,2}}{\tilde P_{N,1} \cdot
\tilde P_{N,2}^2} \cdot {\bf B}_{2} \cdot {\bf g} .
\label{eqn_gradzero}
\end{align}
Rearranging~\eqref{eqn_gradzero} we obtain
\begin{align}
\left( {\bf A}_{1} + \rho_{\rm sig} \cdot {\bf A}_{2} \right)
\cdot {\bf g} = \frac{\tilde P_{R,1}}{\tilde P_{N,1}} \cdot \left(
{\bf B}_{1} + \rho_{\rm noi} \cdot {\bf B}_{2}\right) \cdot {\bf
g} \label{eqn_geneig}
\end{align}
where $\rho_{\rm sig}$ and $\rho_{\rm noi}$ are defined via
\begin{align}
\rho_{\rm sig} = \frac{\tilde P_{R,1}}{\tilde P_{R,2}} \quad
\mbox{and} \quad \rho_{\rm noi} = \frac{\tilde P_{N,1}}{\tilde
P_{N,2}}.
\end{align}

It follows from~\eqref{eqn_geneig} that the optimal ${\bf g}$ must
be a generalized eigenvector of the pair of matrices $\left( {\bf
A}_{1} + \rho_{\rm sig} \cdot {\bf A}_{2} \right)$ and $\left(
{\bf B}_{1} + \rho_{\rm noi} \cdot {\bf B}_{2} \right)$. Moreover,
the corresponding generalized eigenvalue is given by ${\tilde
P_{R,1}} / {\tilde P_{N,1}}$ which is logarithmically proportional
to the rate of the terminal one $r_1$. Unfortunately, the matrices
$\left( {\bf A}_{1} + \rho_{\rm sig} \cdot {\bf A}_{2} \right)$
and $\left( {\bf B}_{1} + \rho_{\rm noi} \cdot {\bf B}_{2}
\right)$ contain the parameters $\rho_{\rm sig}$ and $\rho_{\rm
noi}$ which also depend on ${\bf g}$ and are hence not known in
advance. Therefore, we still need to optimize over these two
parameters. However, compared to the original problem of finding a
complex-valued $M_R \times M_R$ matrix, optimizing over the two
real-valued scalar parameters is already significantly simpler.
The following subsections show how to simplify this 2-D search
even further.

\subsection{Bounds on the parameters $\rho_{\rm sig}$ and
$\rho_{\rm noi}$} Since both parameters $\rho_{\rm sig}$ and
$\rho_{\rm noi}$ have a physical interpretation, the lower and
upper-bounds for them can be easily found. Such bounds are useful
since they limit the search space that has to be tested. For
instance, $\rho_{\rm noi}$ can be expanded into
\begin{align}
\rho_{\rm noi} = \frac{\tilde P_{N,1}}{\tilde P_{N,2}} = \frac{
{\bf g}^H \cdot {\bf J}_1 \cdot {\bf g} + P_{N,1}}{{\bf g}^H \cdot
{\bf J}_2 \cdot {\bf g} + P_{N,2}}. \label{eqn_rhonoi_exp}
\end{align}
The quadratic forms can be bounded by using the fact that for any
Hermitian matrix ${\bf R}$ we have
\begin{align}
\lambda_{\min} ({\bf R}) \cdot \| {\bf g} \|_2^2 \leq {\bf g}^H
\cdot {\bf R} \cdot {\bf g} \leq \lambda_{\max} ({\bf R}) \cdot \|
{\bf g} \|_2^2
\end{align}
where $\lambda_{\min} ({\bf R})$ and $\lambda_{\max} ({\bf R})$
are the smallest and the largest eigenvalues of ${\bf R}$,
respectively. It follows from~\eqref{eqn_def_ji} that
\begin{align}
\lambda_{\min} ({\bf J}_1) = 0 \quad {\rm and} \quad
\lambda_{\max} ({\bf J}_1) = \lambda_{\max} ({\bf R}_{N,R}) \cdot
(\alpha_i^{(b)})^2 \label{rel1}
\end{align}
where $\alpha_i^{(b)}$ is a short hand notation for $\| {\bf
h}_i^{(b)} \|_2$. Furthermore, in general the following inequity
holds
\begin{align}
\lambda_{\max} ({\bf R}_{N,R}) \leq P_{N,R} \cdot M_R \label{rel2}
\end{align}
which for the case of white noise at the relay boils down to the
following tighter condition $\lambda_{\max} ({\bf R}_{N,R}) =
P_{N,R}$.

The relations \eqref{rel1} and \eqref{rel2} can be used to
bound~\eqref{eqn_rhonoi_exp}. Specifically, an upper-bound for
$\rho_{\rm noi}$ can be found by upper-bounding the enumerator and
lower-bounding the denominator, while the lower-bound can be found
by lower-bounding the enumerator and upper-bounding the
denominator. This yields
\begin{align}
\rho_{\rm noi} & \leq \frac{P_{N,R}}{P_{N,2}} \cdot M_R \cdot (
\alpha_1^{(b)} )^2 \cdot \gamma^2 + \frac{P_{N,1}}{P_{N,2}}
\label{eqn_rhonoibnd_1} \\
\rho_{\rm noi} & \geq \left( \frac{P_{N,R}}{P_{N,1}} \cdot M_R
\cdot ( \alpha_2^{(b)} )^2 \cdot \gamma^2 + \frac{P_{N,2}}
{P_{N,1}} \right)^{-1} \label{eqn_rhonoibnd_2}
\end{align}
where $\gamma^2 = \| {\bf g} \|_2^2$ and $M_R$ can be dropped if
the noise at the relay is white. However, an upper-bound for
$\gamma^2$ is till needed. Due to the relay power constraint we
have ${\bf g}^H \cdot {\bf Q} \cdot {\bf g} = P_{T,R}$. Using the
latter condition, the following bound can be derived $\gamma^2
\leq P_{T,R} / \lambda_{\min} ({\bf Q})$. However, it is easy to
check that this bound is very loose since for white noise at the
relay we have $\lambda_{\min} ({\bf Q}) = P_{N,R}$ and for
arbitrary relay noise covariance matrices no lower-bound exists
(the infimum over $\lambda_{\min}$ is zero). This bound is so
loose because it is extremely pessimistic: it measures the norm of
${\bf g}$ in the case when only noise is amplified and no power is
put on the eigenvalues related to the signals of interest.
However, such a case is practically irrelevant since it
corresponds to a sum-rate equal to zero. Therefore, we propose to
replace $\gamma^2$ in~\eqref{eqn_rhonoibnd_1}
and~\eqref{eqn_rhonoibnd_2} by\footnote{We have observed in all
our simulations that this value poses indeed an upper-bound on the
norm of the optimal solution ${\bf g}_{\rm opt}$.}
\begin{align}
\gamma^2 := \frac{( \alpha_1^{(f)} )^2 \cdot ( \alpha_2^{(f)} )^2
\cdot P_{T,1} \cdot P_{T,2} } {( \alpha_1^{(f)} )^2 \cdot P_{T,1}
+ ( \alpha_2^{(f)} )^2 \cdot P_{T,2} }.
\end{align}

In a similar manner, $\rho_{\rm sig}$ can be bounded. In this
case, the enumerator and the denominator have the additional terms
$P_{T,2} \cdot {\bf g}^H \cdot {\bf K}_{2,1} \cdot {\bf g}$ and
$P_{T,1} \cdot {\bf g}^H \cdot {\bf K}_{1,2} \cdot {\bf g}$,
respectively. A pessimistic (loose) bound is obtained by bounding
these two terms independently, i.e., $0 \leq {\bf g}^H \cdot {\bf
K}_{2,1} \cdot {\bf g} \leq \gamma^2 \cdot ( \alpha_2^{(f)} )^2
\cdot ( \alpha_1^{(b)} )^2$ and $0 \leq {\bf g}^H \cdot {\bf
K}_{1,2} \cdot {\bf g} \leq \gamma^2 \cdot (\alpha_1^{(f)})^2
\cdot ( \alpha_2^{(b)} )^2$. This yields
\begin{align}
\rho_{\rm sig} & \leq \frac{P_{T,2}}{P_{N,2}} \cdot
(\alpha_2^{(f)} )^2 \cdot (\alpha_1^{(b)} )^2 \cdot \gamma^2 +
\frac{P_{N,R}}{P_{N,2}} \cdot M_R \cdot (\alpha_1^{(b)} )^2 \cdot
\gamma^2 +
\frac{P_{N,1}}{P_{N,2}} \label{eqn_rhosigbnd_1} \\
\rho_{\rm sig} & \geq \left(\frac{P_{T,1}}{P_{N,1}} \cdot (
\alpha_1^{(f)} )^2 \cdot (\alpha_2^{(b)} )^2 \cdot \gamma^2 +
\frac{P_{N,R}}{P_{N,1}} \cdot M_R \cdot (\alpha_2^{(b)} )^2 \cdot
\gamma^2 + \frac{P_{N,2}}{P_{N,1}} \right)^{-1}.
\label{eqn_rhosigbnd_2}
\end{align}
Again, these bounds are pessimistic since they assume that there
exists an optimal relay strategy for which $P_{R,1} = P_{T,2}
\cdot (\alpha_2^{(f)} )^2 \cdot (\alpha_1^{(b)} )^2 \cdot
\gamma^2$ but $P_{R,2} = 0$, i.e., the rate of the second terminal
is equal to zero. However, it is typically sum-rate optimal to
have significantly more balanced rates between the two users. In
fact, for the ``symmetric'' scenario when $P_{T,1} = P_{T,2}$,
${\bf h}_i^{(f)} = {\bf h}_i^{(b)}$, $i=1, 2$, and $\alpha_1^{(f)}
= \alpha_2^{(f)}$, we always have $P_{R,1} = P_{R,2}$ at the
optimal point.
%\footnote{Unfortunately,
%no analytic proof for this observation could be found up to now.}.
Therefore, these bounds can be further tightened if a priori
knowledge about the scenario is available.

\subsection{Efficient 2-D and 1-D Search}
Once the search space for $\rho_{\rm sig}$ and $\rho_{\rm noi}$
has been fixed, we can find the maximum via optimization over
these two parameters using a 2-D search. In general, a 2-D
exhaustive search can be computationally demanding, i.e., the
complexity will be higher than that of the POTDC algorithm.
However, as we show in the sequel, for the problem at hand, this
search can be implemented efficiently. These efficient
implementations are, however, heuristic since they rely on
properties of the objective functions that are apparent by visual
inspection. As we will see in simulations, the resulting RAGES
algorithm performs as well as the rigorous POTDC algorithm in
practice.

Fig.~\ref{fig_costfcn_ex} demonstrates a typical example of the
sum-rate $r_1+r_2$ as a function of $\rho_{\rm sig}$ and
$\rho_{\rm noi}$. For this example we have chosen $M_R = 6$,
$P_{T,1} = P_{T,2} = P_{T,R} = 1$, $P_{N,1} = P_{N,2} = P_{N,R} =
0.1$ and we have drawn the channel vectors from an uncorrelated
Rayleigh fading distribution assuming reciprocity. By visual
inspection, this sample objective function shows two interesting
properties. First, it is a quasi-convex function with respect to
the parameters $\rho_{\rm noi}$ and $\rho_{\rm sig}$ which allows
for efficient (quasi-convex) optimization tools for finding its
maximum. Albeit this property is only demonstrated for one example
here, it has been always present in our numerical evaluations even
when largely varying all system parameters. Secondly, for every
value of $\rho_{\rm sig}$ the corresponding maximization over
$\rho_{\rm noi}$ yields one maximal value which depends on
$\rho_{\rm noi}$ only very weakly. This is illustrated by
Fig.~\ref{fig_maxvsrhonoi_ex} which displays the relative change
of the objective function $r_1+r_2$ for different choices of
$\rho_{\rm noi}$, each time optimizing it over $\rho_{\rm sig}$.
The displayed values represent the relative decrease of the
objective functions compared to the global optimum, i.e., for the
worst choice of $\rho_{\rm noi}$, the achieved sum-rate is about
$2 \cdot 10^{-5} = 0.002$~\% lower than for the best choice of
$\rho_{\rm noi}$. Consequently, the 2-D search over $\rho_{\rm
sig}$ and $\rho_{\rm noi}$ can be replaced essentially without any
loss by a 1-D search over $\rho_{\rm sig}$ only for one fixed
value of $\rho_{\rm noi}$ (e.g., the geometric mean of the upper
and the lower-bound).

In addition, instead of performing the search directly over the
original objective function $r_1 + r_2$, we can find an even
simpler objective functions by using the physical meaning of our
two search parameters. To this end, let us introduce a new
parameter $\hat{\rho}_{\rm sig}$ as a function of $\bf g$ as
follows
\begin{align}
\hat{\rho}_{\rm sig} ({\bf g}) = \frac{{\bf g}^H \cdot {\bf A}_1
\cdot {\bf g}}{{\bf g}^H \cdot {\bf A}_2 \cdot {\bf g}} .
\end{align}
Here ${\bf g}$ is the relay weight vector at the current search
point ($\rho_{\rm sig}, \rho_{\rm noi}$). Then we know that in the
optimal point ${\bf g}_{\rm opt}$, we have $\hat{\rho}_{\rm
sig}({\bf g}_{\rm opt}) = {\rho}_{\rm sig}$. This can be used to
construct a new objective function
\begin{align}
A_{\rm sig}(\rho_{\rm sig}, \rho_{\rm noi}) = \hat{\rho}_{\rm
sig}({\bf g}_{\rm opt}) - {\rho}_{\rm sig}
\end{align}

Using the same data set as before, we display the corresponding
shape of $A_{\rm sig}(\rho_{\rm sig}, \rho_{\rm noi})$ in
Fig.~\ref{fig_asig_ex}. The red dashed line indicates the set of
points where $A_{\rm sig}(\rho_{\rm sig}, \rho_{\rm noi}) = 0$. It
can be observed that for every value of $\rho_{\rm noi}$, $A_{\rm
sig}(\rho_{\rm sig}, \rho_{\rm noi})$ is a monotonic function in
$\rho_{\rm sig}$. Therefore, the bisection method can be used to
find a zero crossing in $\rho_{\rm sig}$ which coincides with the
sum-rate-optimal $\rho_{\rm sig}$ for a given $\rho_{\rm noi}$.

\subsection{Summary}
In summary, it can be concluded that the RAGES approach simplifies
the optimization over a complex-valued $M_R \times M_R$ matrix
into the optimization over two real-valued parameters which both
have a physical interpretation. Even more, the 2-D search can be
simplified into a 1-D search by fixing one of the parameters. The
loss incurred to this step is typically small. In the example
provided above, it is only 0.002~\%, but even varying the system
parameters largely and using many random trials we never found a
relative difference higher than a few percents.

Moreover, the 1-D search can be efficiently implemented by
exploiting the quasi-convexity of $r_1 + r_2$ %(e.g., using a branch-and-bound algorithm)
or the monotonicity of $A_{\rm sig}$ (e.g., using the bisection
method). Again, these properties are only demonstrated by examples
but we have observed in all our simulations that the resulting algorithm
yields a sum-rate very close to the optimum found by the exact
solution and its upper-bound described before. This comparison is
further illustrated in next section via numerical simulations.

Comparing the POTDC and RAGES approaches, it is noticeable that the
POTDC approach is absolutely rigorous, while the RAGES approach is
at some points heuristic. The complexity of solving the proposed
sum-rate maximization problem for two-way AF MIMO relaying using
the POTDC algorithm is the same as the complexity of solving the
semi-definite programming problem \eqref{final_relaxed} and
iterating over a single parameter $\beta$. The typical number of
iterations is 4-7. Alternatively, the complexity of solving the
same problem using the RAGES approach is equivalent to the
complexity of finding the dominant generalized eigenvector, which
has to be performed for each combination of the parameters
$\rho_{\rm sig}$ and $\rho_{\rm noi}$. Since, as has been shown,
the search over one parameter only is sufficient, the complexity
of the RAGES approach is typically lower than that
of the POTDC algorithm, especially for the 1-D RAGES.

\section{Simulation Results}
In this section, we evaluate the performance of the new proposed
methods via numerical simulations. Consider a communication system
consisting of two single-antenna terminals and an AF MIMO relay
with $M_R$ antenna elements. The communication between the
terminals is bidirectional, i.e., it is performed based on the
two-way relaying scheme. It is assumed that perfect channel
knowledge is available at the terminals and at the relay, while
the terminals use only effective channels (scalars), but the relay
needs full channel vectors. The relay estimates the corresponding
channel coefficients between the relay antenna elements and the
terminals based on the pilots which are transmitted from the
terminals. Then based on these channel vectors, the relay computes
the relay amplification matrix $\mathbf G$ and then uses it for
forwarding the pilot signals to the terminals. After receiving the
forwarded pilot signals from the relay via the effective channels,
the terminals can estimate the effective channels using a suitable
pilot-based channel estimation scheme, e.g., the LS.

The noise powers of the relays and the terminals $P_{N,R}$,
$P_{N,1}$ and $P_{N,2}$ are assumed to be equal to $\sigma^{2}$.
Uncorrelated Rayleigh fading channels are considered and it is
assumed that reciprocity holds, i.e., ${\bf h}_i^{(f)} = {\bf
h}_i^{(b)}$ for $i=1,2$. The relay is assumed to be located on a
line of unit length which connects the terminals to each other and
the variances of the channel coefficients between terminal $i,
i=1,2$ and the relay antenna elements are all assumed to be
proportional to $1/d_i^\nu$, where $ d_i \in (0,1)$ is the
normalized distance between the relay and the terminal $i$ and
$\nu$ is the path-loss exponent which is assumed to be equal to
$3$ throughout the simulations. \footnote{It is experimentally
found that typically $2 \leq \nu \leq 6$ (see
\cite[p.~46--48]{Goldsmith} and references therein). However,
$\nu$ can be smaller than 2 when we have a wave-guide effect,
i.e., indoors in corridors or in urban scenarios with narrow
street canyons.} For obtaining each simulated point, $100$
independent simulation runs are used unless otherwise is
specified.

In order to design the relay amplification matrix $\mathbf G$,
five different methods are considered including the proposed
POTDC, 2-D RAGES and 1-D RAGES algorithms, the algebraic
norm-maximizing (ANOMAX) transmit strategy of \cite{ANOMAX} and
the discrete Fourier transform (DFT) method that chooses the relay
precoding matrix as a scaled DFT matrix.  Note that the ANOMAX
strategy provides a closed-form solution for the problem. Also
note that for the DFT method no channel knowledge is needed. Thus,
the DFT method serves as a benchmark for evaluating the gain
achieved by using channel knowledge. The upper-bound is also shown
in all simulations. For obtaining the upper-bound, the interval
$[\beta_{\min}, \beta_{\max}]$ is divided in $30$ segments. In
addition, the proposed techniques are compared to the
SNR-balancing technique of \cite{Shahram} for the relevant to the
later technique scenario when multiple single-antenna relay nodes
are used.

\subsection{Example~1: Symmetric Channel Conditions} In our first
example, we consider the case when the channels between the relay
antenna elements and both terminals have the same channel quality.
More specifically, it is assumed that the relay is located in the
middle of the connecting line between the terminals and the
transmit power of the terminals $P_{T,1}$ and $P_{T,2}$ and the
total transmit power of the MIMO relay $P_{T,R}$ are all assumed
to be equal to $1$.

{Fig.~\ref{3relaysSCSP1} shows the sum-rate achieved by different
aforementioned methods versus $\sigma^{-2}$ for the case of
$M_R=3$. It can be seen in this figure that the performance of the
proposed methods coincides with the upper-bound. Thus, the methods
perform optimally in terms of providing the maximum sum-rate. The
ANOMAX technique performs close to the optimal, while the DFT
method gives a significantly lower sum-rate.

\subsection{Example~2: Asymmetric Channel Conditions}
In the second example, we consider the case when the channels
between the relay antenna elements and the second terminal have
better channel quality than the channels between the relay antenna
elements and the first terminal and, and evaluate the effect of
the relay location  on the achievable sum-rate. Particularly, we
consider the case when the distance between the relay and the
second terminal, $d_2$, is less than or equal to the distance
between the relay and the first terminal, $d_1$. The total
transmit power of the terminals, i.e., $P_{T,1}$ and $P_{T,2}$ and
the total transmit power of the MIMO relay $P_{T,R}$ all are
assumed to be equal to $1$ and the noise power in the relays and
the terminals all are assumed to be equal to $1$.

Fig.~\ref{3relaysASCASP1} shows the sum-rate achieved in this
scenario by different aforementioned methods versus the distance
between the relay and the second terminal denoted as $d_2$, for
the case of $M_R=3$. It can be seen in this figure that the
proposed methods perform optimally, while the performance
(sum-rate) of ANOMAX is slightly worse.

As mentioned earlier, it is guaranteed that the POTDC algorithm
converges to at least a local maximum of the sum-rate maximization
problem. However, our extensive simulation results confirm that
the POTDC algorithm converges to the global maximum of the problem
in all simulation runs. Indeed, the performance of the POTDC
algorithm coincides with the upper-bound. Moreover, the 2-D RAGES
and 1-D RAGES are, in fact, globally optimal, too. The ANOMAX and
DFT methods, however, do not achieve the maximum sum-rate. The
loss in sum-rate related to the DFT method is quite significant
while the loss in sum-rate related to the ANOMAX method grows from
small in the case of symmetric channel conditions to significant
in the case of asymmetric channel conditions. Although ANOMAX
enjoys a closed-form solution and it is even applicable in the
case when terminals have multiple antennas, it is not a good
substitute for the proposed methods for the sum-rate maximization
goal, because of this significant gap in performance in the
asymmetric case.

\subsection{Example~3: Effect of The Number of Relay Antenna Elements}
In this example, we consider the effect of the number of relay
antenna elements $M_R$ on the achievable sum-rate for the
aforementioned methods. The powers assigned to the first and the
second terminals as well as to the relay are all equal to 1. The
relay is assumed to be located at the distance of $1/4$ from the
second user. Moreover, the noise powers at the terminals and at
the relay antenna elements are all assumed to be equal to $1$. For
obtaining each simulated point in this simulation example, $200$
independent simulation runs are used.

Fig.~\ref{SUMrateVERSUSMR} depicts the sum-rates achieved by
different methods versus the number of relay antenna elements
$M_R$. As it is expected, by increasing $M_R$ (thus, increasing
the number of degrees of freedom), the sum-rate increases. For the
DFT method, the sum rate does not increase with an increase in the
number of the relay antennas because of the lack of channel
knowledge for this method. The proposed methods achieve higher
sum-rate compared to ANOMAX.

\subsection{Example~4: Performance Comparison for the Scenario
of Two-Way Relaying via Multiple Single-Antenna Relays} In our
last example, we compare the proposed methods with the SNR
balancing-based approach of \cite{Shahram}. The method of
\cite{Shahram} is developed for a two-way relaying system which
consists of two single-antenna terminals and multiple
single-antenna relay nodes. Subject to the constraint on the total
transmit power of the relay nodes and the terminals, the method of
\cite{Shahram} designs a beamforming vector for the relay nodes
and the transmit powers of the terminals to maximize the minimum
received SNR at the terminals. In order to make a fair comparison,
we consider a diagonal structure for the relay amplification
matrix $\mathbf G$ that corresponds to the special case of
\cite{Shahram} when multiple single-antenna nodes are used for
relaying. It is worth mentioning that for imposing such a diagonal
structure for the relay amplification matrix $\mathbf G$ in POTDC
and RAGES, the vector $\mathbf g_{M_R^2 \times 1} = {\rm
vec}(\mathbf G)$ is replaced with $\mathbf g_{M_R \times 1} ={\rm
diag} (\mathbf G)$ and the matrices $\mathbf A_i$ and $\mathbf
B_i, i=1,2$ are replaced with new square matrices $\mathbf
{\tilde{A}}_i$ and $\mathbf {\tilde{B}}_i, i=1,2$ of size $M_R
\times M_R$ such that $\mathbf {\tilde{A}}_i\ (m,n)= \mathbf A_i \
((m-1) \cdot M_R+m,(n-1) \cdot M_R+n)$ and $\mathbf {\tilde{B}}_i\
(m,n)= \mathbf B_i \ ((m-1) \cdot M_R+m,(n-1) \cdot M_R+n), \ \
m,n=1,\cdots,M_R$. Moreover, we assume fixed transmit powers at
the terminals and choose them to be equal to $1$. The total
transmit power at the relay also equals $1$ and the relay is
assumed to lie in the middle of the terminals. Fig.~\ref{Shahram}
shows the corresponding performance of the different methods. From
this figure it can be observed that the proposed methods
demonstrate a significantly better performance compared to the
method of \cite{Shahram} as it may be expected.

\section{Conclusions and Discussions}
We have shown that the sum-rate maximization problem in two-way AF
MIMO relaying belongs to the class of DC programming problems.
Although the typical approach for solving the DC programming
problems is the branch-and-bound method, it does not have any
polynomial time guarantees for its worst-case complexity.
Therefore, we have developed in this paper two algorithms for
finding the global maximum of the aforementioned problem with
polynomial time worst-case complexity. The POTDC algorithm is
based on a specific parameterization of the objective function,
that is, the product of quadratic ratios, and then application of
semi-definite programming (SDP) relaxation, linearization and
iterative search over a single parameter. Its design is rigorous
and is based on the recent advances in convex optimization. To the
best of our knowledge, this is the first polynomial time algorithm
for solving a class of DC programming problems rigorously. The
RAGES algorithm is based on a different parameterization of the
objective function and the generalized eigenvectors method, but
may enjoy a lower computational complexity that makes it a valid
alternative especially if 1-D search is used. The upper-bound for
the solution of the problem is developed and it is demonstrated by
simulations that both proposed method achieve the upper-bound and
are, thus, globally optimal.

The proposed POTDC algorithm represents a general optimization
technique applicable for solving a wide class of DC programming
problems. Essentially, the optimization problems consisting of the
maximization/minimization of a product of quadratic ratios can be
handled using the proposed POTDC approach. Moreover, the POTDC
algorithm can be used for solving the optimization problems
containing in any of the constraints a difference of two quadratic
forms. Some relatively straightforward modifications may, however,
be required. For example, if the problem is to optimize a product
of more than two quadratic ratios under a single quadratic (power)
constraint, the number of constraints in the corresponding DC
programming problem will be more than three. Thus, the result used
in this paper that even after relaxing the rank-one constraint in
the step of SDP relaxation, it is possible to find algebraically
an exact rank-one solution based on the solution of the relaxed
problem, does not hold any longer. Then, randomization procedures
will have to be adopted to recover a rank-one solution from the
solution of the relaxed problem. In this case, such solutions
obviously may not be exact, but all the results related to the SDP
relaxation will apply.

Other signal processing problems that can be addressed using the
proposed POTDC approach are the general-rank robust adaptive
beamformer with a positive semi-definite constraint
\cite{ArashMe}, the dynamic spectrum management for digital
subscriber lines \cite{LeNgoc}, the problems of finding the
weighted sum-rate point, the proportional-fairness operating point
and the max-min optimal point  for MISO interference channel
\cite{JL10}, the problem of robust beamforming design for AF relay
networks with multiple relay nodes and so on. The extensions of
the POTDC approach to some of the aforementioned problem is an
issue of future research.

\newpage
\begin{figure}[t]
    \psfrag{AA}[cc][cc]{\Large \textsf{UT$_1$}}
    \psfrag{BB}[cc][cc]{\Large \textsf{UT$_2$}}
    \psfrag{C}[cc][cc]{\Large \textsf{RS}}
      \psfrag{hh}{\small $1$}
    \psfrag{kk}{\small $M_R$}
    \center{
    \includegraphics[width=60mm, height=35mm, trim=12mm 12mm 12mm 5mm]{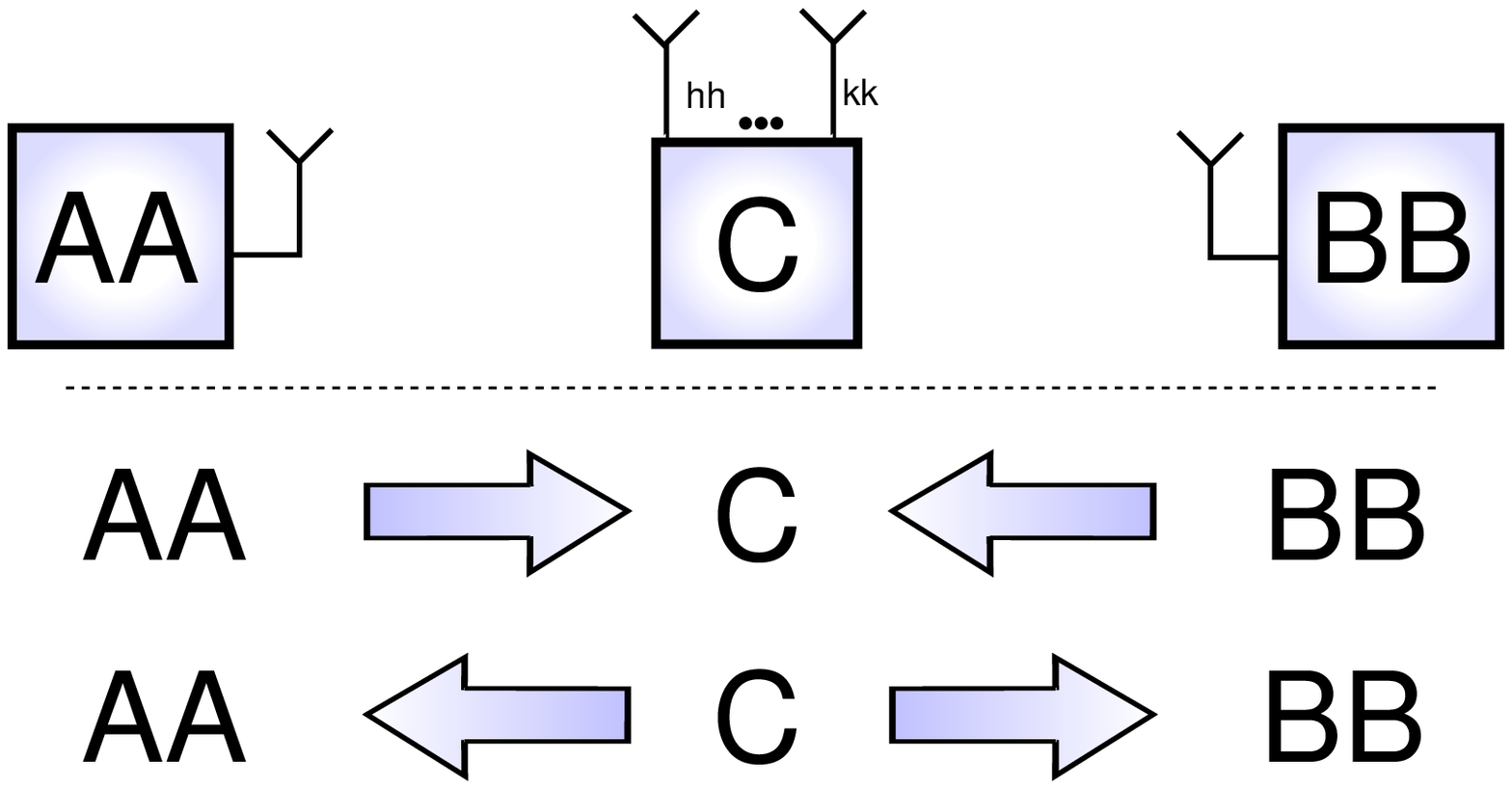}
    }
   \caption{Two-way relaying system model.}
   \label{fig_system}
\end{figure}

\begin{figure}[t]
\begin{center}
\includegraphics[scale=.75]{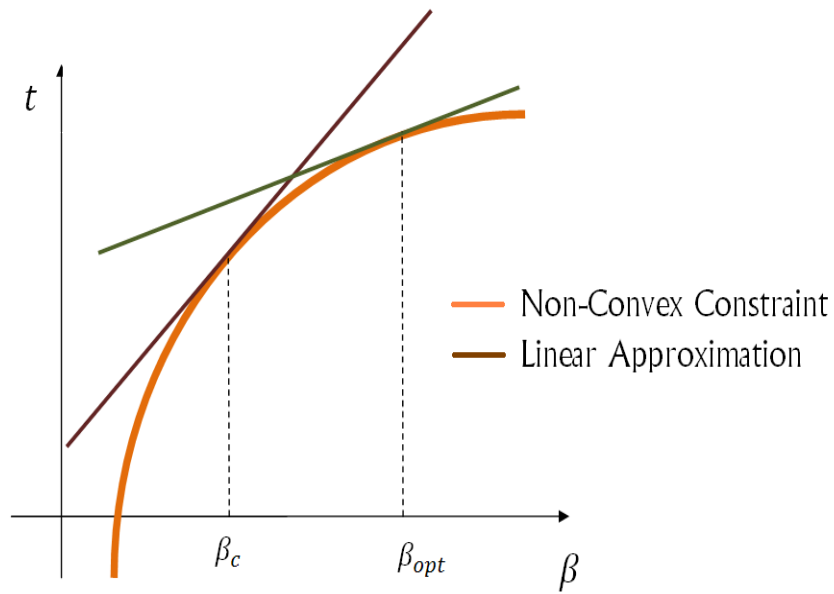}
\caption{Linear approximation of $\log(\beta)$ around $\beta_c$.}
\label{linearization}
\end{center}
\end{figure}

\begin{figure}[t]
\begin{center}
\includegraphics[scale=.75]{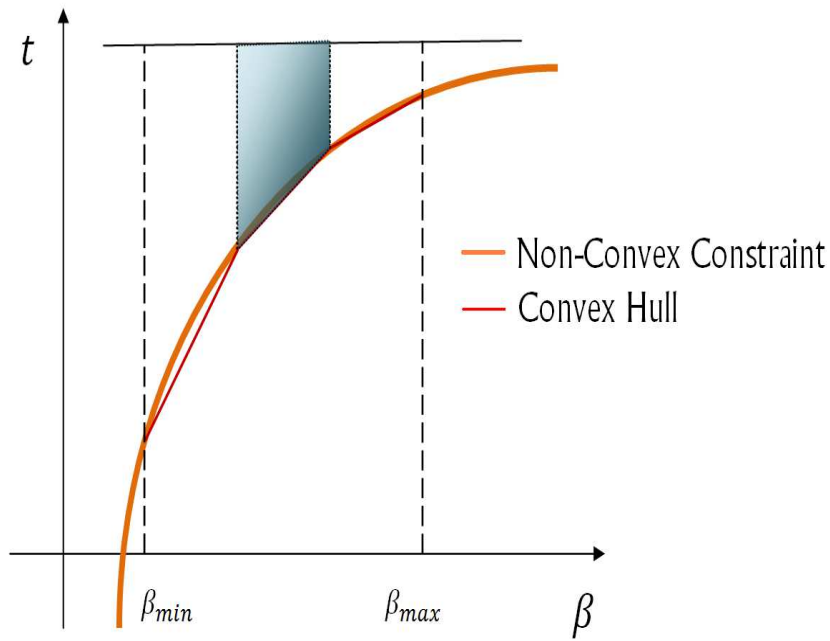}
\caption{Feasible region of the constraint $\log(\beta) \leq t$
and the convex hull in each sub-division.} \label{Fig1}
\end{center}
\end{figure}

\begin{figure}[t]%
\begin{center}
\includegraphics[scale=.58]{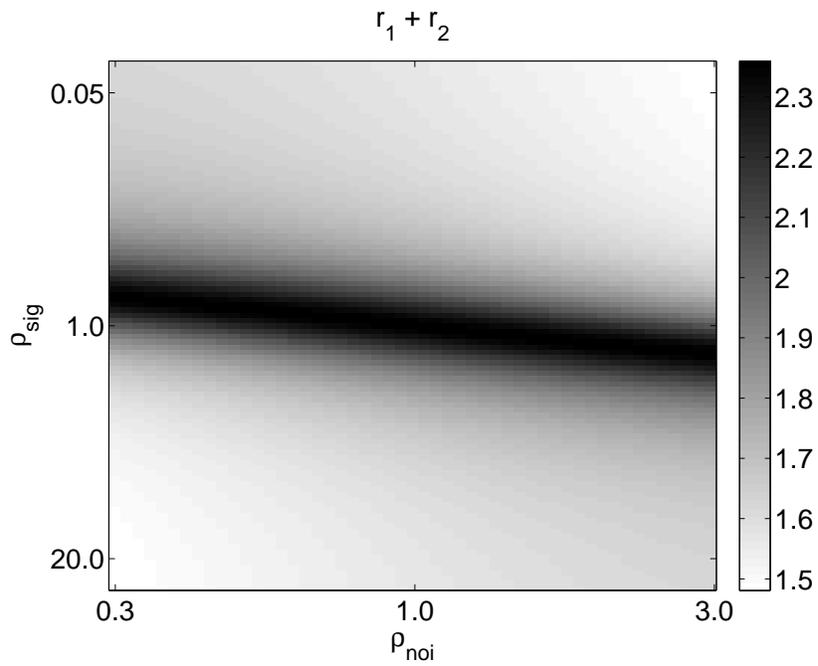}%
\caption{Sum-rate $r_1+r_2$ versus $\rho_{\rm sig}$ and $\rho_{\rm
noi}$ for $M_R=6$, $P_{T,1} = P_{T,2} = P_{T,R} = 1$, $P_{N,1} =
P_{N,2} =
P_{N,R} = 0.1$.}%
\label{fig_costfcn_ex}%
\end{center}
\end{figure}

\begin{figure}[t]%\
\begin{center}
\includegraphics[scale=.58]{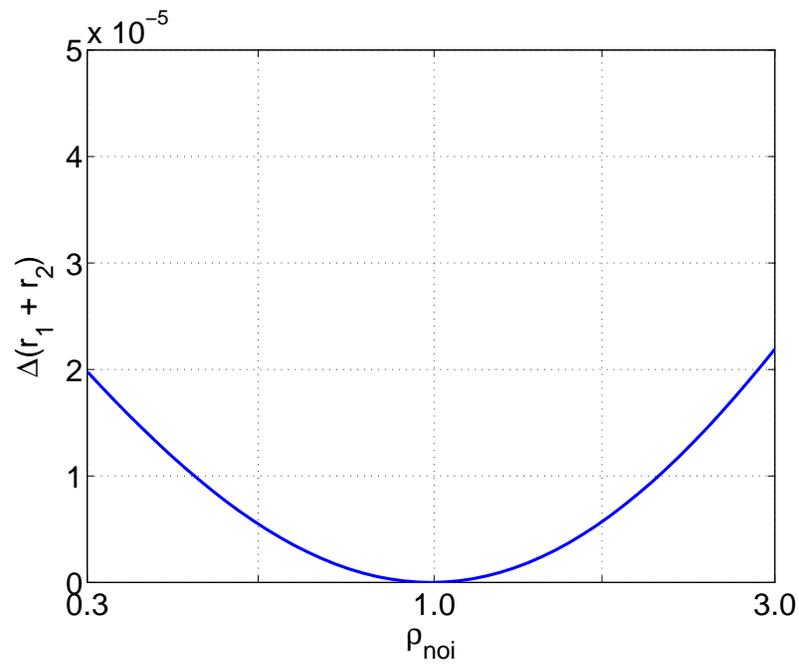}%
\caption{Relative change in sum-rate $r_1+r_2$ versus $\rho_{\rm
noi}$: optimizing over $\rho_{\rm sig}$ for every choice of
$\rho_{\rm noi}$. The
same data set as in Fig.~\ref{fig_costfcn_ex} is used.}%
\label{fig_maxvsrhonoi_ex}%
\end{center}
\end{figure}

\begin{figure}[t]%
\begin{center}
\includegraphics[scale=.58]{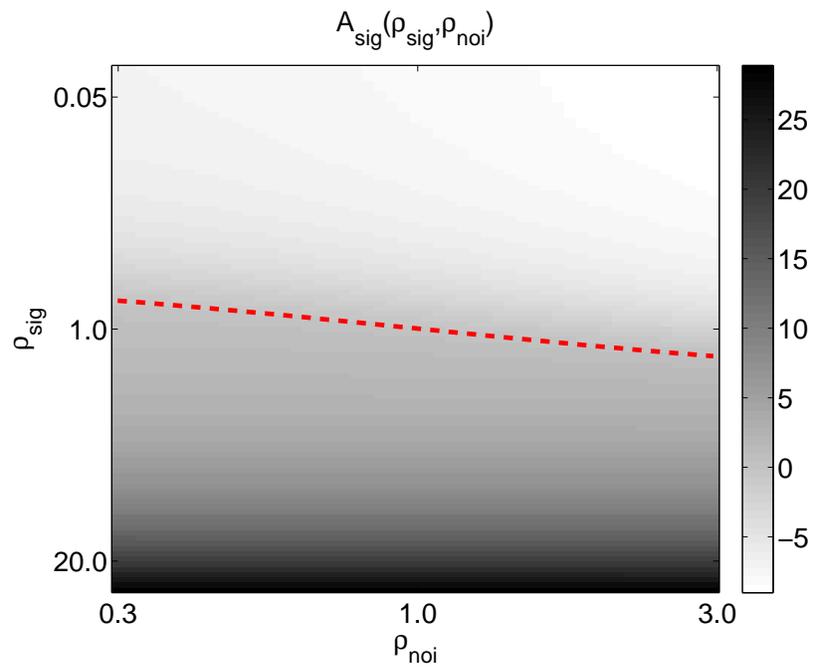}%
\caption{Objective function $A_{{\rm sig}}(\rho_{\rm
sig},\rho_{\rm noi})$. The same data set as in
Fig.~\ref{fig_costfcn_ex} is used. The red dashed
line indicates the points where $A_{{\rm sig}}(\rho_{\rm sig},
\rho_{\rm noi}) = 0$.}%
\label{fig_asig_ex}%
\end{center}
\end{figure}

\begin{figure}[t]
\begin{center}
\includegraphics[scale=.49]{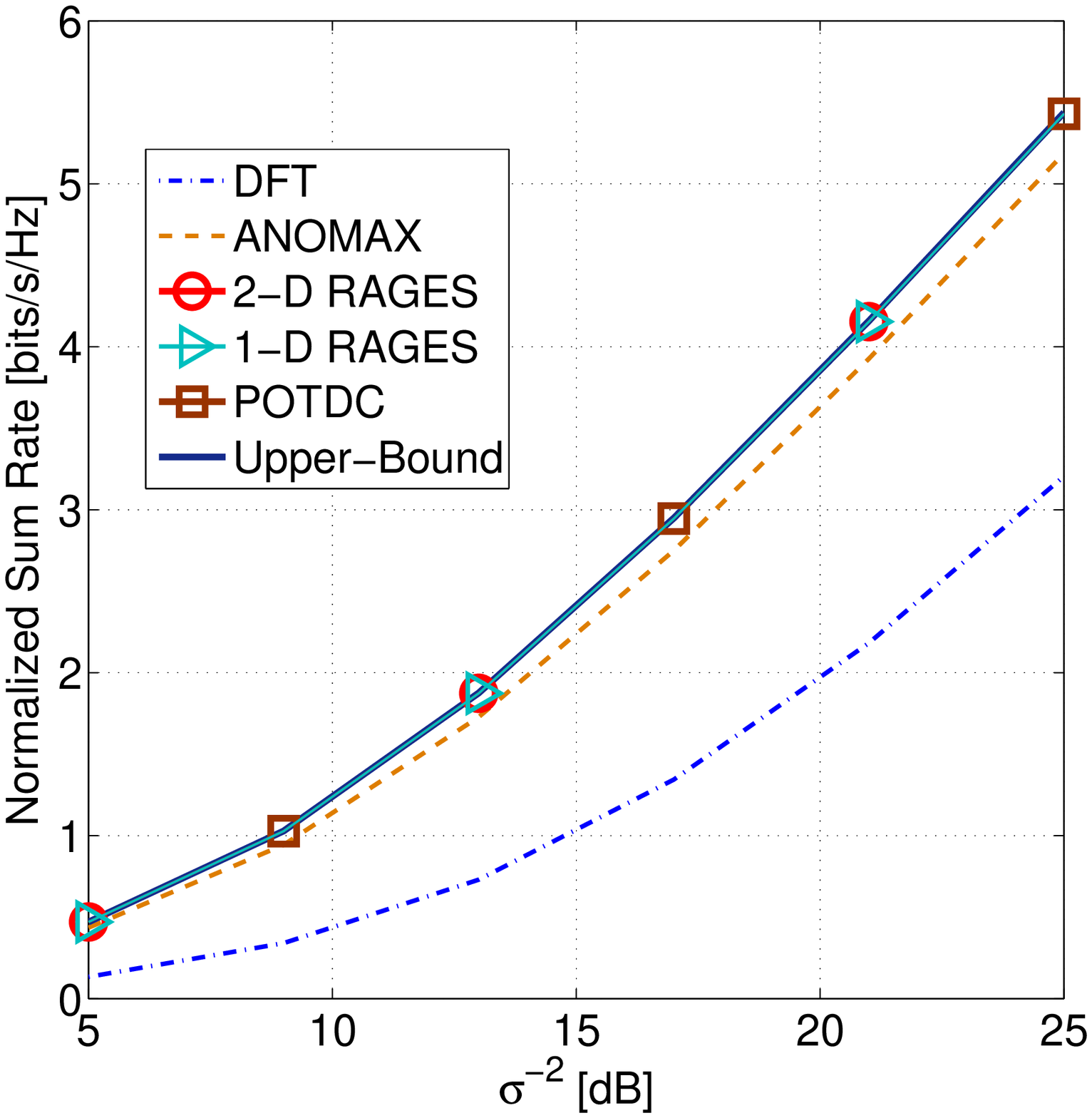}
\caption{Sum-rate versus $\sigma^{-2}$ for $M_R=3$ antennas. The
case of symmetric channel conditions.} \label{3relaysSCSP1}
\end{center}
\end{figure}

\begin{figure}[t]
\begin{center}
\includegraphics[scale=.49]{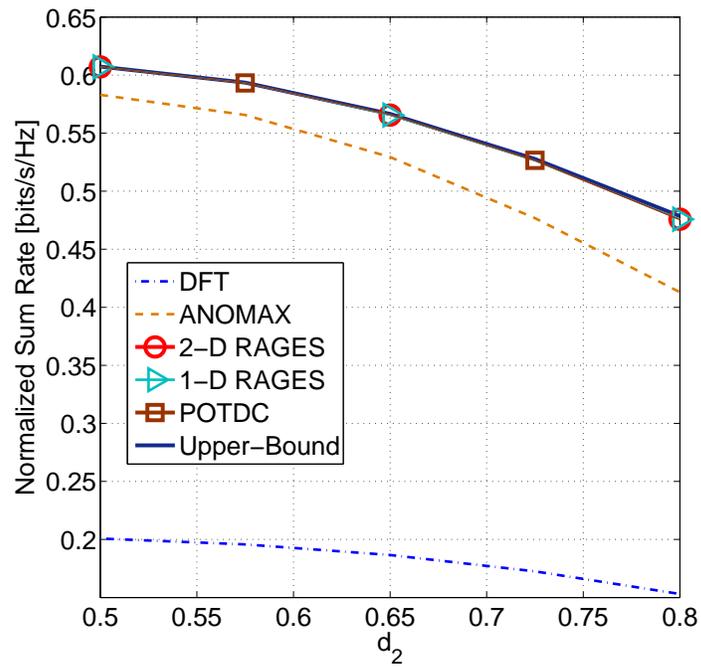}
\caption{Sum-rate versus the distance between the relay and the
second terminal $d_2$ for $M_R=3$ antennas. The case of asymmetric
channel conditions.} \label{3relaysASCASP1}
\end{center}
\end{figure}

\begin{figure}[t]
\begin{center}
\includegraphics[scale=.49]{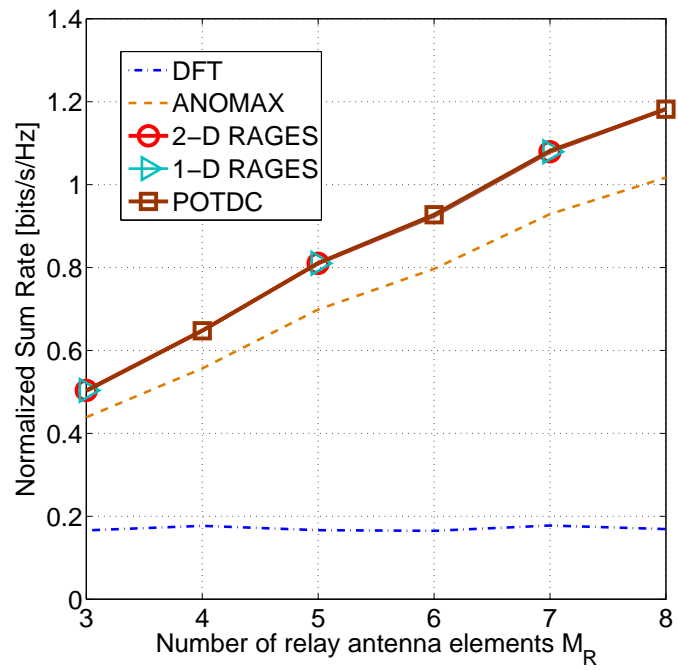}
\caption{Sum-rate versus the number of the relay antenna elements
$M_R$. The case of asymmetric channel conditions.}
\label{SUMrateVERSUSMR}
\end{center}
\end{figure}

\begin{figure}[t]
\begin{center}
\includegraphics[scale=.49]{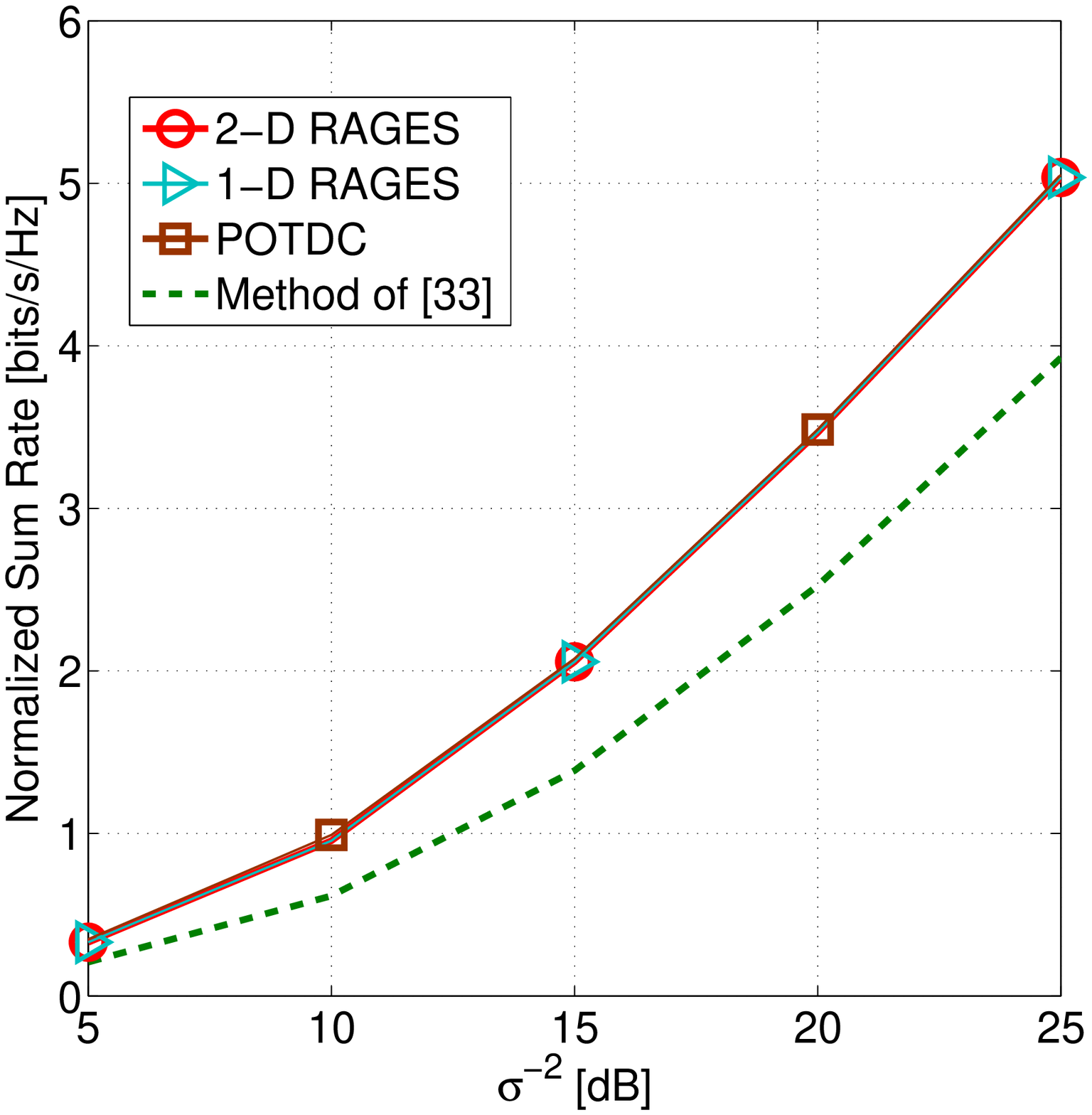}
\caption{Sum-rate versus $\sigma ^{-2}$ for $M_R=3$ antennas. The
case of symmetric channel conditions.} \label{Shahram}
\end{center}
\end{figure}

\end{document}